\documentclass[12pt]{iopart}
\pdfoutput=1
\usepackage{graphicx}
\usepackage{siunitx}
\usepackage{multirow}

\usepackage{array}
\newcolumntype{L}{>{$}l<{$}} 

\begin{document}

\title{Precise sinusoidal signal extraction from noisy waveform in vibration calibration}
\author[T. Shimoda {\it et al.}]{Tomofumi Shimoda,  Wataru Kokuyama, Hideaki Nozato}

\address{National Metrology Institute of Japan, National Institute of Advanced Industrial Science and Technology, 1-1-1 Umezono, Tsukuba, Ibaraki 305-8563, Japan}
\ead{tomofumi.shimoda@aist.go.jp}
\vspace{10pt}

\begin{abstract}
Precise extraction of sinusoidal vibration parameters is essential for the dynamic calibration of vibration sensors, such as accelerometers.
However, several standard methods have not yet been optimized for large background noise.
In this work, signal processing methods to extract small vibration signals from noisy data in the case of accelerometer calibration is discussed.
The results show that spectral leakage degrades calibration accuracy.
Three methods based on the use of a filter, window function, and numerical differentiation are investigated with theoretical calculations, simulations, and experiments.
These methods can effectively reduce the contribution of the calibration system noise.
The uncertainty of micro vibration calibration in the National Metrology Institute of Japan is reduced by two orders of magnitudes using the proposed methods.
The theoretical analyses in this work can lay the foundation for the optimization of signal processing in vibration calibration, and can be applied to other dynamic calibration fields.
\end{abstract}

\section{Introduction} \label{sec:introduction}
Micro vibration measurement is required in various fields, such as infrastructure health monitoring \cite{ISHM} or satellite performance analysis \cite{ESA.handbook}.
Many types of sensors have been developed and used, such as high-sensitivity accelerometers, broadband seismometers, or low-noise microelectromechanical system (MEMS) accelerometers (e.g., \cite{MEMS.Deng,MEMS.Isobe,MEMS.review}). 
The calibration of the sensor frequency response is essential for the reliability of micro vibration measurements.
As the applications of micro vibration measurement increase, measuring responses to small input vibrations is becoming increasingly important.

Accurate extraction of amplitude and phase from a sinusoidal waveform is required for the calibration of accelerometer sensitivity.
The target accelerometer is sinusoidally vibrated by a vibration exciter, and the amplitudes and phases of the sensor voltage signal and reference displacement signal are estimated and compared to calibrate the sensitivity and phase shift of the accelerometer.
The estiation accuracy is essential for the calibration uncertainty.
The requirement for the primary calibration is about 0.1~\% and 0.1$^\circ$ for the amplitude and phase, respectively \cite{CCAUV.V-K2,CCAUV.V-K3}.
In these studies, large vibration that was nearly $10$~m/s$^2$ was applied.
To obtain the response to a micro vibration with an amplitude down to about $10^{-3}$~m/s$^2$, the applied vibration amplitude should also be small (on the same order of magnitude) because the linearity of the response is not ensured in general.
The extraction of such a small vibration signal usually suffers from the background noise of the calibration system, which originates from the background vibration or electrical noise.

For accelerometer calibration, amplitude and phase extraction are performed by the sine approximation method (SAM) in ISO16063-11 \cite{ISO16063-11}.
However, the SAM is not optimized for real cases with large background noise.
In addition to the background noise reduction of the calibration system, the optimization of signal processing is important. 
Some of the calibration institutes empirically apply a digital filter to deal with the problem, although an unified approach has not yet been established.
Another signal processing method using correlation has also been proposed \cite{COPA}.
In this paper, we discuss the limitations of the SAM and its optimization through theoretical investigations, simulations, and experiments. 
These investigations are important to calibrate accelerometers under a large noise or with a small vibration amplitude.
The results can be applied to not only accelerometer calibration but also other signal processing settings that require accurate sinusoidal parameter estimation.

The remainder of this paper is organized as follows: Section~\ref{sec:calculation} summarizes the mathematical background of the SAM and its verification with simulations. 
Section~\ref{sec:reduce} proposes optimization methods, including filtering, changing the window function, and numerical differentiation. 
Section~\ref{sec:experiment} applies the proposed method to the vibration calibration in the National Metrology Institute of Japan (NMIJ).

\section{Contribution of system noise to sinusoidal signal extraction} \label{sec:calculation}
\subsection{Mathematical framework of the conventional SAM}\label{sec:framework}
\begin{figure}
	\begin{center}
	\includegraphics[width=10cm]{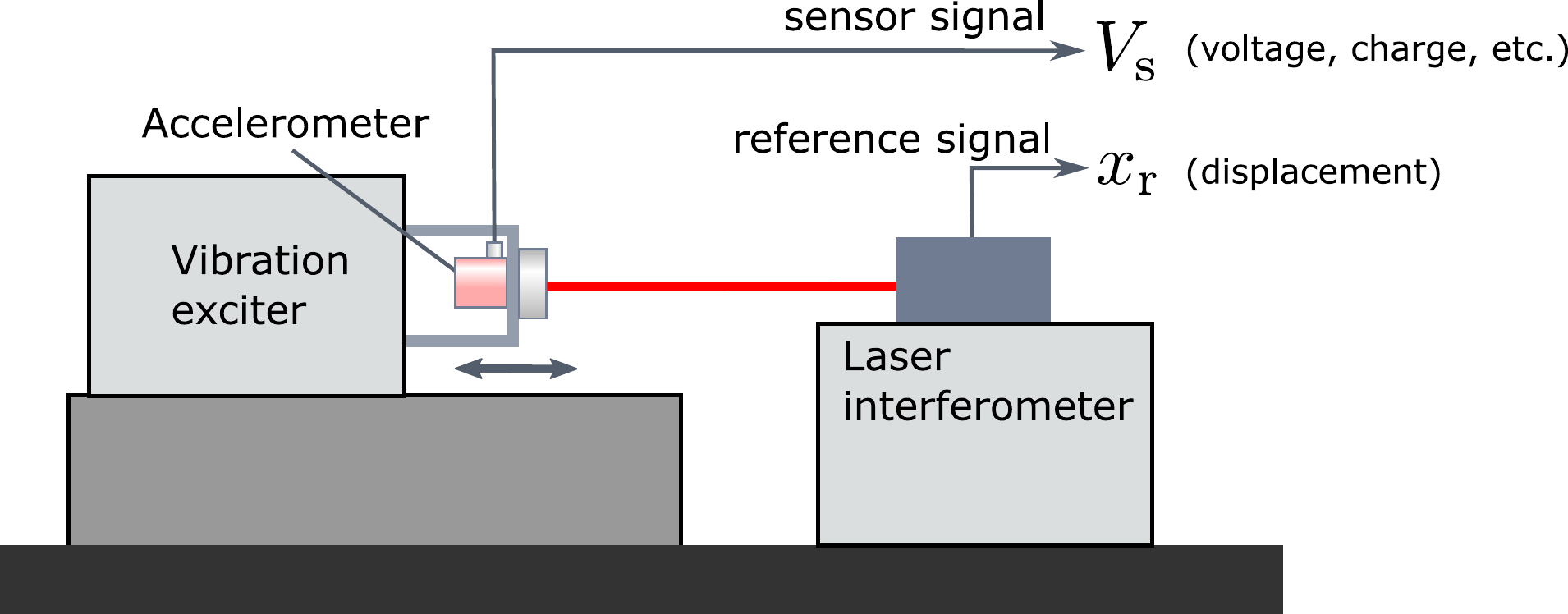}
	\caption{Primary calibration system of an accelerometer.}
	\label{fig:system}
	\end{center}
\end{figure}
The overview of the accelerometer calibration system is depicted in Fig.~\ref{fig:system}.
The accelerometer is sinusoidally vibrated by the vibration exciter, and its output signal $V_\mathrm{s}$ is recorded along with the reference displacement signal $x_\mathrm{r}$ measured by the laser interferometer.
Throughout this paper, signal processing is assumed to be digital signal processing using discretely sampled data.
The SAM specified in ISO16063-11 \cite{ISO16063-11} determines the sensitivity modulus and phase shift from two waveforms.
The amplitude and phase of each waveform at vibration frequency $f_\mathrm{v}$ are extracted in the following process.
Here, a waveform $x(t)$ is sampled within $0<t<T$ at $t_n=nT/N$ ($n=$0, 1, ..., $N-1$) ($N$: the number of data).
The recorded data $x_n$ is modeled as 
\begin{equation}
	x_n = b_0 + b_1 \cos(2\pi f_\mathrm{v} t_n) + b_2 \sin(2\pi f_\mathrm{v} t_n) + \epsilon_n,
\end{equation}
where $b_0$, $b_1$, and $b_2$ are the fitting parameters, and $\epsilon_n$ is the residual from the model.
In this work, we assume that the sampling is sufficiently faster than the vibration frequency, $N/T \gg f_\mathrm{v}$, and the data length is an integer multiple of the vibration period, $T=N_\mathrm{c}/f_\mathrm{v}$ ($N_\mathrm{c}$: integer).
These conditions are easily achieved using a commercially available digitizer and by setting the proper record length.
Under these conditions, the optimal parameters that minimize the residual $\sum\epsilon_n^2$ are given by
\begin{equation}
	b_1 = \frac{2}{N} \sum_{n=0}^{N-1} x_n \cos(2\pi f_\mathrm{v} t_n), \ \ b_2 = \frac{2}{N} \sum_{n=0}^{N-1} x_n \sin(2\pi f_\mathrm{v} t_n), 	\label{eq:b1b2}
\end{equation}
which are identical to the Fourier transform of the data.
Here, the estimated complex amplitude for the variable $x$ is defined as $\hat{x}_\mathrm{est}\equiv b_1+ib_2$ for convenience.
The amplitudes are computed for both the sensor signal $V_\mathrm{s}$ and the reference signal $x_\mathrm{r}$ to calculate the sensitivity modulus and phase delay as 
\begin{equation}
	S_\mathrm{cal}=\frac{|\hat{V}_\mathrm{s,est}|}{(2\pi f_v)^2 |\hat{x}_\mathrm{r,est}|}, \hspace{20pt}
	\Delta\phi_\mathrm{cal} = \arg(\hat{V}_\mathrm{s,est}) - \arg(\hat{x}_\mathrm{r,est}) - \pi ,
	\label{eq:Scal}
\end{equation}
respectively.

Using Eq.~(\ref{eq:b1b2}), the complex amplitude $\hat{x}_\mathrm{est}$ can be modified as
\begin{eqnarray}
	\hat{x}_\mathrm{est}(t_0) &=& \frac{2}{N} \sum_{n=0}^{N-1} x_n e^{2\pi i f_\mathrm{v} t_n} \nonumber\\
	&\simeq& \frac{2}{T} \int_0^T x(t) e^{2\pi i f_\mathrm{v} t} dt \nonumber \\
	&=& \frac{2}{T} \int_{-\infty}^\infty w(t+t_0) x(t) e^{2\pi i f_\mathrm{v} t} dt \nonumber \\
	&=& \frac{2}{T} \int_{-\infty}^\infty \tilde{W}(f) \tilde{X}(f_\mathrm{v}-f) e^{-2\pi i f t_0} df \label{eq:c}
\end{eqnarray}
Here, $t_0$ is the start time of the measurement, and
\begin{equation}
	w(t) = w_\mathrm{r}(t) = \left\{ \begin{array}{@{\kern2.5pt}lL}
	1\ (0<t<T) \\
	0\ (\rm{otherwise})\end{array}
	\right.
\end{equation}
is a rectangular window function. 
$\tilde{W}(f)$ and $\tilde{X}(f)$ are the Fourier transform of $w(t)$ and $x(t)$, respectively.
In summary, the conventional SAM is the Fourier transformation with the rectangular window function.

As Eq.~(\ref{eq:c}) indicates, the parameter estimation at $f_\mathrm{v}$ is affected by the frequency component $X(f)\ (f\neq f_\mathrm{v})$, which is known as the spectral leakage.
For the rectangular window, the leakage is determined by 
\begin{equation}
	\tilde{W}_\mathrm{r}(f) = T \mathrm{sinc}(fT) e^{\pi if T}, \label{eq:Wr}
\end{equation}
where sinc function is defined as $\mathrm{sinc}(z)=\sin(\pi z)/\pi z$.
Although the amplitude of the purely sinusoidal vibration $x_0(t)=\hat{x}_0\sin(2\pi f_\mathrm{v} t)$ is calculated to be $\hat{x}_{0\mathrm{,est}}(f_\mathrm{v})= i \hat{x}_0$, the background noise at $f_\mathrm{v}$ or the leakage from $f\neq f_\mathrm{v}$ can be a problem in real-world cases.

\subsection{SAM under noise}\label{sec:SAM.noise}
\begin{figure}
	\begin{center}
	\includegraphics[width=15cm]{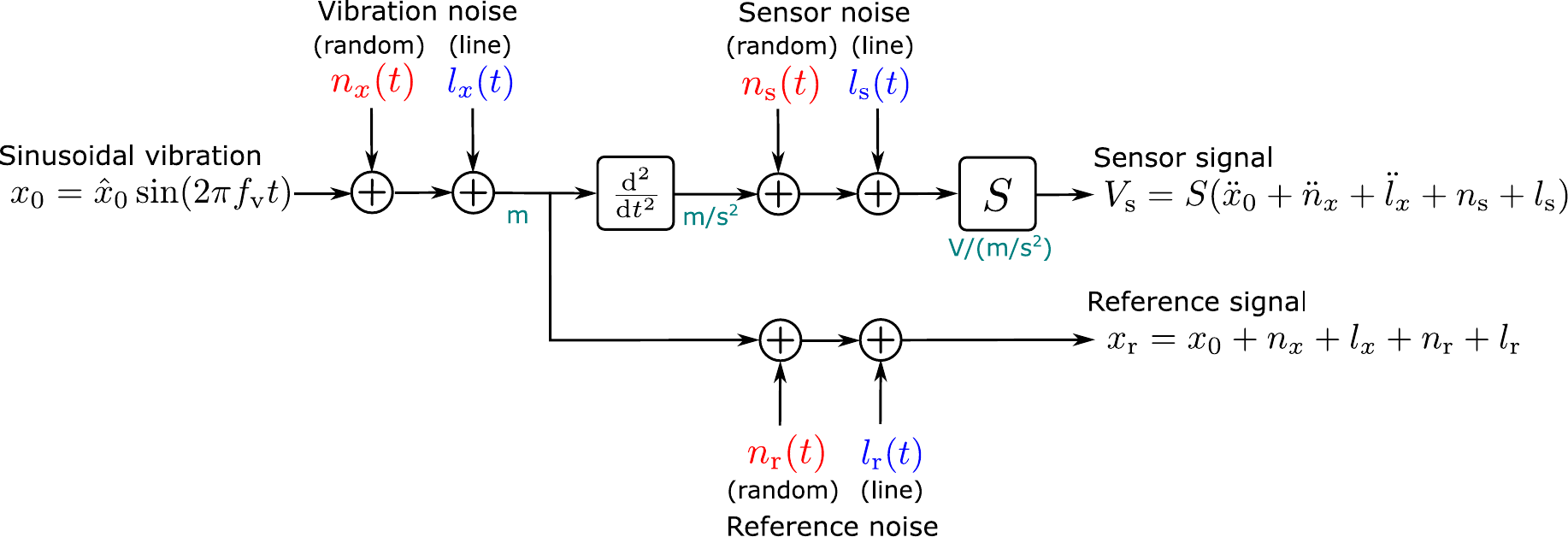}
	\caption{Model of the signal flow in accelerometer calibration. }
	\label{fig:model}
	\end{center}
\end{figure}
Fig.~\ref{fig:model} shows the signal flow model for accelerometer calibration shown in Fig.~\ref{fig:system}.
A similar discussion is applicable to other experiments conducted to measure the amplitude ratio between two vibration timeseries.
The error sources considered in Fig.~\ref{fig:model} are divided into four types 
\begin{itemize}
\item{independent random background noise of each device ($n_\mathrm{s}$, $n_\mathrm{r}$)}
\item{common random noise to both sensor and reference signals ($n_x$)}
\item{independent line noise or harmonics of each device ($l_\mathrm{s}$, $l_\mathrm{r}$)}
\item{common line noise or harmonics to both sensor and reference signals ($l_x$)}
\end{itemize}
In the calibration process, sinusoidal vibration $x_0=\hat{x}_0\sin(2\pi f_\mathrm{v}t)$ is applied.
The random vibration noise $n_x$ and line noise $l_x$ are added due to the background vibration, electrical noise of the system, or the distortion of the waveform.
Here, $l_x$ includes the line noise, such as the power supply noise, which appears at constant frequencies, and the harmonics of the input vibration, which appear at $2f_\mathrm{v}$, $3f_\mathrm{v}$, and so on.
Then, the reference signal measures the displacement of the waveform, while the sensor to be calibrated responds to the second derivative of it.
Each output signal contains the independent random and line noise; $n_\mathrm{r}$, $l_\mathrm{r}$, $n_\mathrm{s}$, and $l_\mathrm{s}$.
The sensor outputs the signal with the sensitivity modulus of $S$.
The recorded signals are the sum of these contributions, as shown in Fig.~\ref{fig:model}.
The calculated amplitudes of the recorded signals $\hat{V}_\mathrm{s,est}$ and $\hat{x}_\mathrm{r,est}$ are affected by the noise components as
\begin{eqnarray}
	\hat{V}_\mathrm{s,est} = S ( \hat{\ddot{x}}_\mathrm{0,est} + \hat{\ddot{n}}_{x\mathrm{,est}} + \hat{\ddot{l}}_{x\mathrm{,est}} + \hat{n}_\mathrm{s,est} + \hat{l}_\mathrm{s,est} ) \\
	\hat{x}_\mathrm{r,est} = \hat{x}_{0\mathrm{,est}} + \hat{n}_{x\mathrm{,est}} + \hat{l}_{x\mathrm{,est}} + \hat{n}_\mathrm{r,est} + \hat{l}_\mathrm{r,est} ,
\end{eqnarray}
which results in $S_\mathrm{cal}\neq S$ due to the second or later terms.
Note that the harmonics generated through a nonlinear process, such as the nonlinearity of the sensor, are not independent of the input signal; for example, $l_\mathrm{s}$ can be correlated to $l_x$ in general.
However, we assumed that the error sources in Fig.~\ref{fig:model} were independent of each other for simplicity.
The random noise is discussed in Section~\ref{sec:SAM.random}, and the line noise and harmonics are discussed in Section~\ref{sec:SAM.line}.

\subsection{Effect of random noise} \label{sec:SAM.random}
We consider the general case for random noise $n(t)$, which is characterized with (one-sided) power spectral density (PSD) $G(f)$.
The standard deviation of the estimated amplitude for the noise, $\sqrt{\langle |\hat{n}_\mathrm{est}|^2 \rangle}$, is the standard uncertainty of the amplitude of $x(t)=x_0(t)+n(t)$.
From Eq.~(\ref{eq:c}), the real and imaginary parts of $\hat{n}_\mathrm{est}(t_0)$ can be written as
\begin{eqnarray}
	\mathrm{Re}[\hat{n}_\mathrm{est}(t_0)] &=& \frac{1}{T} \int_{-\infty}^\infty \tilde{W}(f) \left\{ \tilde{N}(f_\mathrm{v}-f) + \tilde{N}^\ast(f_\mathrm{v}+f) \right\} e^{-2\pi i f t_0} df, \label{eq:Re_n} \\
	\mathrm{Im}[\hat{n}_\mathrm{est}(t_0)] &=& \frac{1}{iT} \int_{-\infty}^\infty \tilde{W}(f) \left\{ \tilde{N}(f_\mathrm{v}-f) - \tilde{N}^\ast(f_\mathrm{v}+f) \right\} e^{-2\pi i f t_0} df \label{eq:Im_n}	.
\end{eqnarray}
$\tilde{N}(f)$ is the Fourier spectrum of $n(t)$.
The PSD corresponding to the Fourier spectrum $\tilde{N}(f_\mathrm{v}-f) \pm \tilde{N}^\ast(f_\mathrm{v}+f)$ is given by $G(|f_\mathrm{v}-f|) + G(|f_\mathrm{v}+f|)$.
Therefore, the standard deviations of Eq.~(\ref{eq:Re_n}) and (\ref{eq:Im_n}) are both given by the integral of the PSD $|\tilde{W}(f)|^2 (G(|f_\mathrm{v}-f|) + G(|f_\mathrm{v}+f|)) / T^2$.
Since $\mathrm{Re}[\hat{n}_\mathrm{est}(t_0)]$ and $\mathrm{Im}[\hat{n}_\mathrm{est}(t_0)]$ are independent from each other because of the randomness of $n(t)$, the amplitude estimation uncertainty under the random noise is derived as
\begin{eqnarray}
	\sqrt{\langle |\hat{n}_\mathrm{est}|^2 \rangle} &=& \sqrt{ \langle \mathrm{Re}[\hat{n}_\mathrm{est}]^2 \rangle } =  \sqrt{ \langle \mathrm{Im}[\hat{n}_\mathrm{est}]^2 \rangle } \\
	&=& \sqrt{ \int_0^\infty \frac{|\tilde{W}(f)|^2}{T^2} \left\{ G(|f_\mathrm{v}-f|) + G(|f_\mathrm{v}+f|) \right\} df }\\
	&=& \sqrt{ \int_{-\infty}^\infty \frac{|\tilde{W}(f_\mathrm{v}-f)|^2}{T^2} G(|f|) df } \label{eq:std.n}
\end{eqnarray}
This equation represents the uncertainty of vibration amplitude estimation under the random noise with the PSD of $G(f)$.
The uncertainty follows a Gaussian distribution.

Note that the phase estimation uncertainty (in radian) under the random noise is identical to the relative uncertainty of the amplitude: 
\begin{equation}
	u(\arg[\hat{x}_\mathrm{est}]) 
	= u \left(\arctan\left( \frac{\mathrm{Im}[\hat{x}_0+\hat{n}_\mathrm{est}]}{\mathrm{Re}[\hat{x}_0+\hat{n}_\mathrm{est}]} \right) \right) 
	= \frac{\sqrt{\langle |\hat{n}_\mathrm{est}|^2 \rangle}}{\hat{x}_0}
	= \frac{u(|\hat{x}_\mathrm{est}|)}{\hat{x}_0}.
	\label{eq:std.arg}
\end{equation}
This is because the distribution of $\hat{n}_\mathrm{est}(t_0)$ is isotropic in the complex plane.
The component with the same argument as $\hat{x}_0$ is the amplitude error, and the orthogonal component is the phase error.
Therefore, we investigated only the amplitude uncertainty using Eq.~(\ref{eq:std.n}) for the random noise in the following subsections and Section~\ref{sec:reduce} for simplicity as the same results are applicable to the phase.
Their equivalence is confirmed in Section~\ref{sec:experiment}.

\subsubsection{Independent random noise for the sensor and reference\\} \label{sec:indep.random}
The sensor and reference signals contain independent random background noise $n_\mathrm{s}$ and $n_\mathrm{r}$.
They contribute to the measurement result via $\hat{n}_\mathrm{s,est}$ and $\hat{n}_\mathrm{r,est}$, which have random values for different measurements.
Their standard deviations are the standard uncertainty of the amplitude $V_\mathrm{s}$ and $x_\mathrm{r}$.
Their expressions $\langle |\hat{n}_\mathrm{s,est}|^2 \rangle$ and $\langle |\hat{n}_\mathrm{r,est}|^2 \rangle$ are given by Eq.~(\ref{eq:std.n}) using the PSDs $G_\mathrm{s}(f)$ and $G_\mathrm{r}(f)$.
Consequently, the relative standard uncertainty of the calibration sensitivity in the absence of the other noise sources is 
\begin{eqnarray}
	\hspace{-10mm}
	\frac{u(S_\mathrm{cal})}{S} 
	&=& \sqrt{ \left( \frac{u(|\hat{V}_\mathrm{s,est}|)}{S|\hat{\ddot{x}}_\mathrm{0,est}|} \right)^2 + \left( \frac{u(|\hat{x}_\mathrm{r,est}|)}{|\hat{x}_{0\mathrm{,est}}|} \right)^2 }	  \\
	&=& \frac{1}{(2\pi f_\mathrm{v})^2\hat{x}_0} \sqrt{ \int_{-\infty}^\infty \frac{|\tilde{W}(f_\mathrm{v}-f)|^2}{T^2} \left\{ G_\mathrm{s}(|f|) + (2\pi f_\mathrm{v})^4 G_\mathrm{r}(|f|) \right\} df }.
\label{eq:uScal.random}
\end{eqnarray}

In the limit of long measurement time $T$, the window function is asymptotically identical to the Dirac delta function as $\lim_{T \to \infty}|\tilde{W}(f)|^2=T\delta(f)$; hence, the amplitude estimation uncertainty is 
\begin{equation}
	\sqrt{ \langle |\hat{n}_\mathrm{s,est}|^2 \rangle} \rightarrow \sqrt{ \frac{G_\mathrm{s}(f_\mathrm{v})}{T} }. \label{eq:std.cns.approx}
\end{equation}
This relation indicates that the uncertainty is determined by the noise spectrum at the vibration frequency and is inversely proportional to the square-root of the measurement time.
Eq.~(\ref{eq:std.cns.approx}) gives the theoretical limit of measurements, which cannot be avoided unless the background noise of the system is reduced.

In reality, spectral leakage happens as shown in Eq.~(\ref{eq:std.n}).
This means that even if the background noise is small at $f_\mathrm{v}$, the overall signal-to-noise ratio (S/N) can be degraded by the noise in the other frequency band.
The amount of spectral leakage is determined by the shape of the window function.
Therefore, the leakage can be reduced by changing the window function $w(t)$ or filtering the signal $V_\mathrm{s}$ and $x_\mathrm{r}$ to suppress the noise at $f\neq f_\mathrm{v}$.
The details of these modifications are discussed in Section~\ref{sec:reduce}.

\begin{figure}
	\begin{minipage}{0.5\hsize}
		\begin{center}
		\includegraphics[width=8cm]{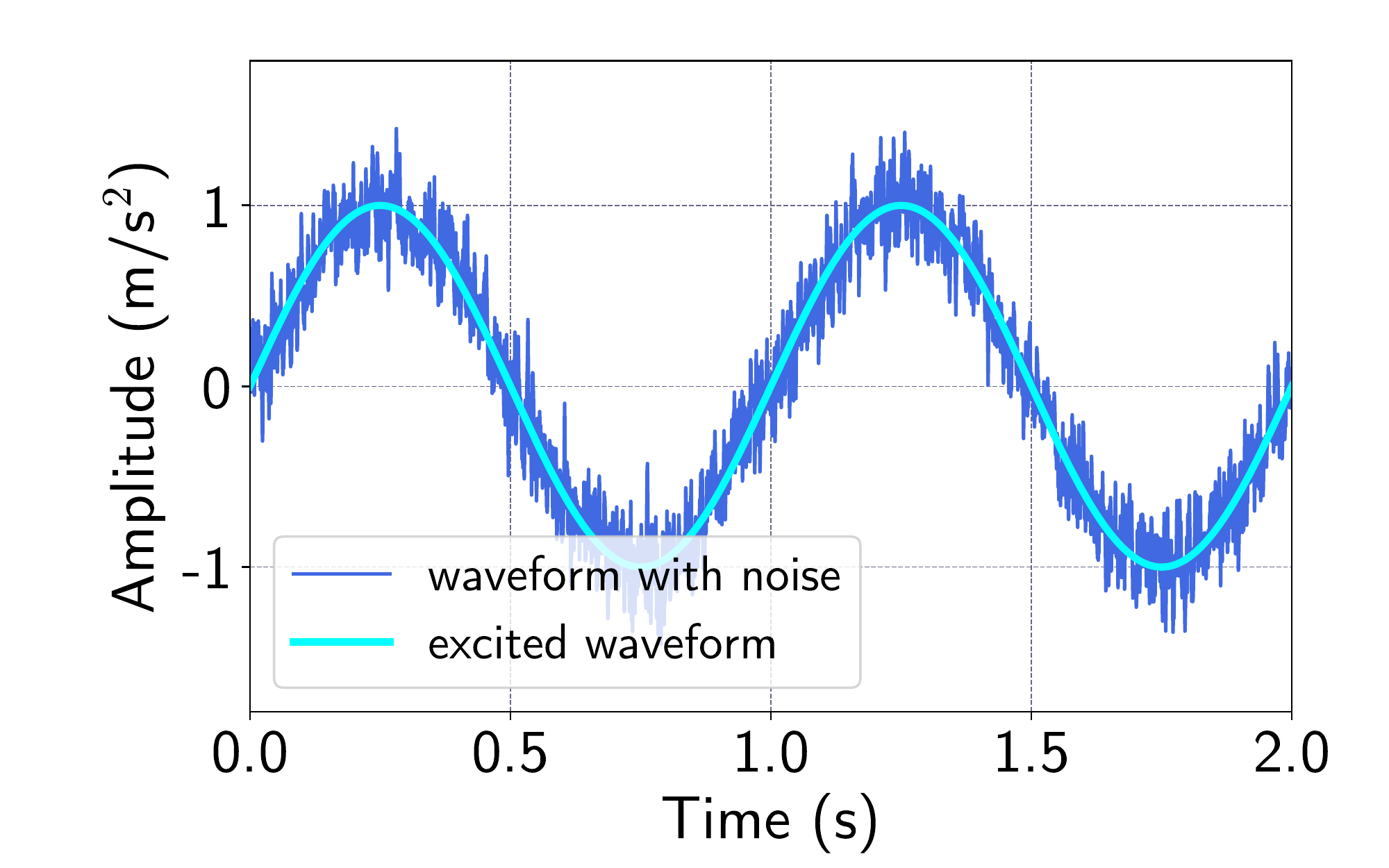}
		\end{center}
	\end{minipage}
	\begin{minipage}{0.5\hsize}
		\begin{center}
		\includegraphics[width=8cm]{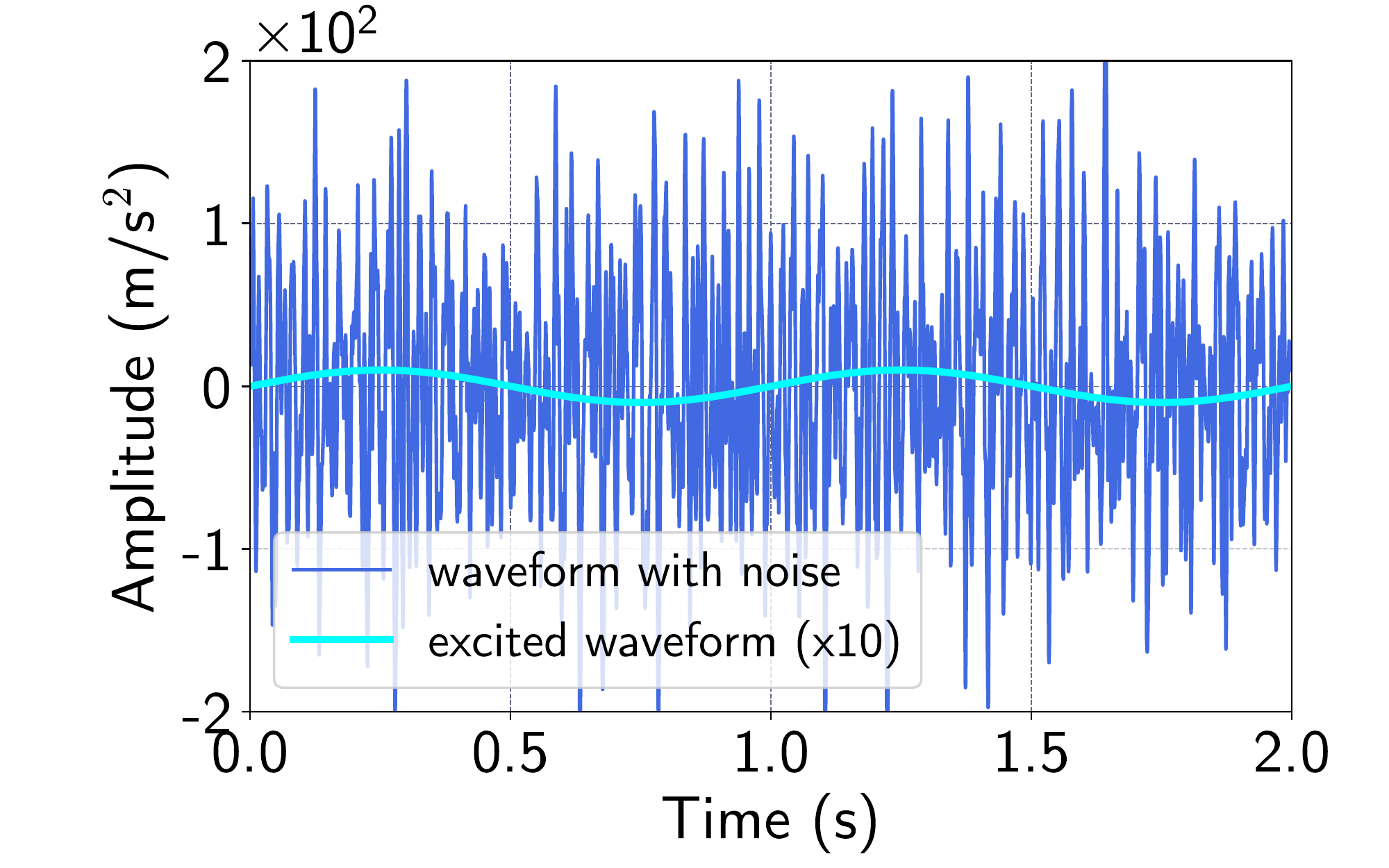}
		\end{center}
	\end{minipage}
	\\
	\begin{minipage}{0.5\hsize}
		\begin{center}
		\includegraphics[width=8cm]{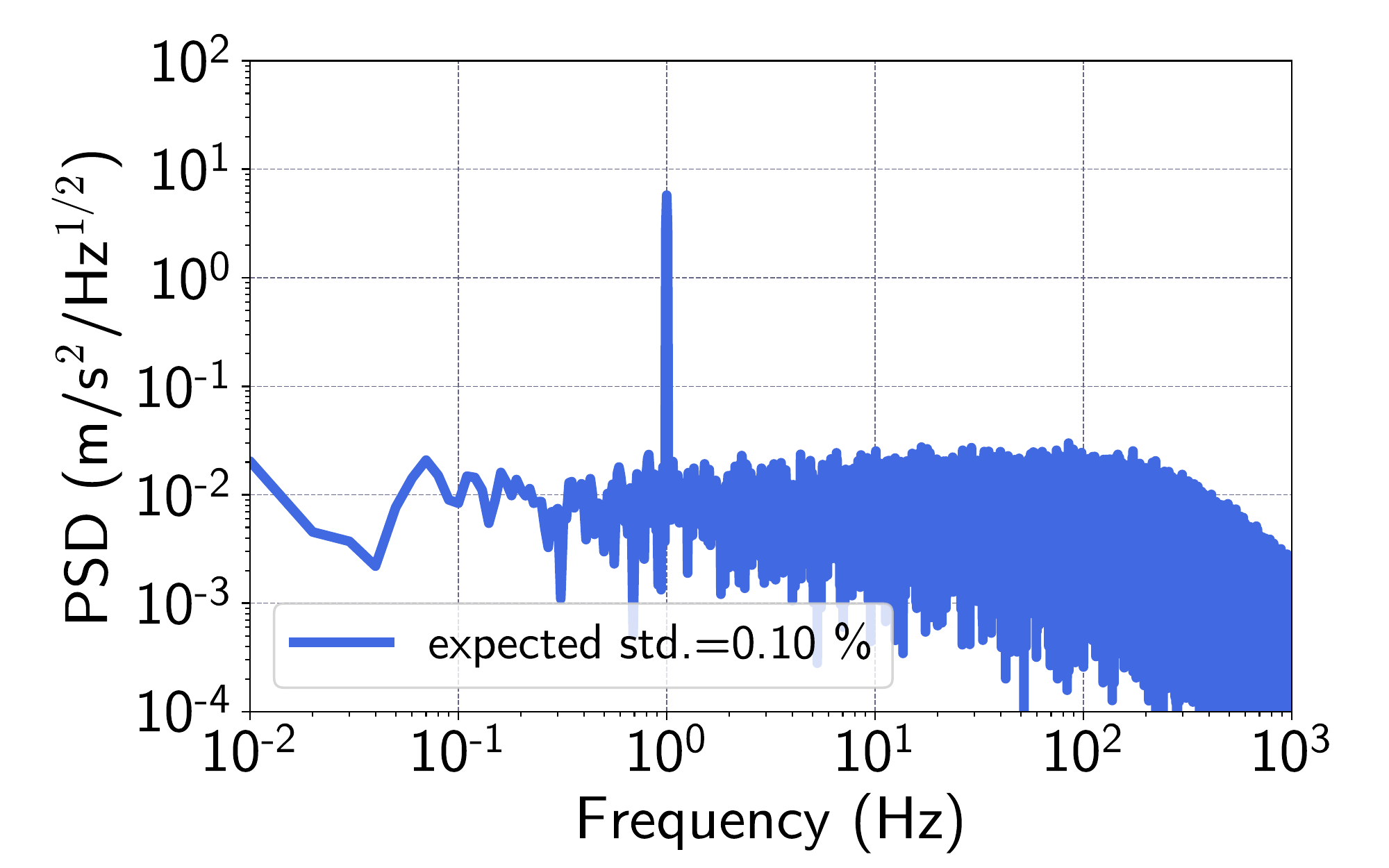}
		\end{center}
	\end{minipage}
	\begin{minipage}{0.5\hsize}
		\begin{center}
		\includegraphics[width=8cm]{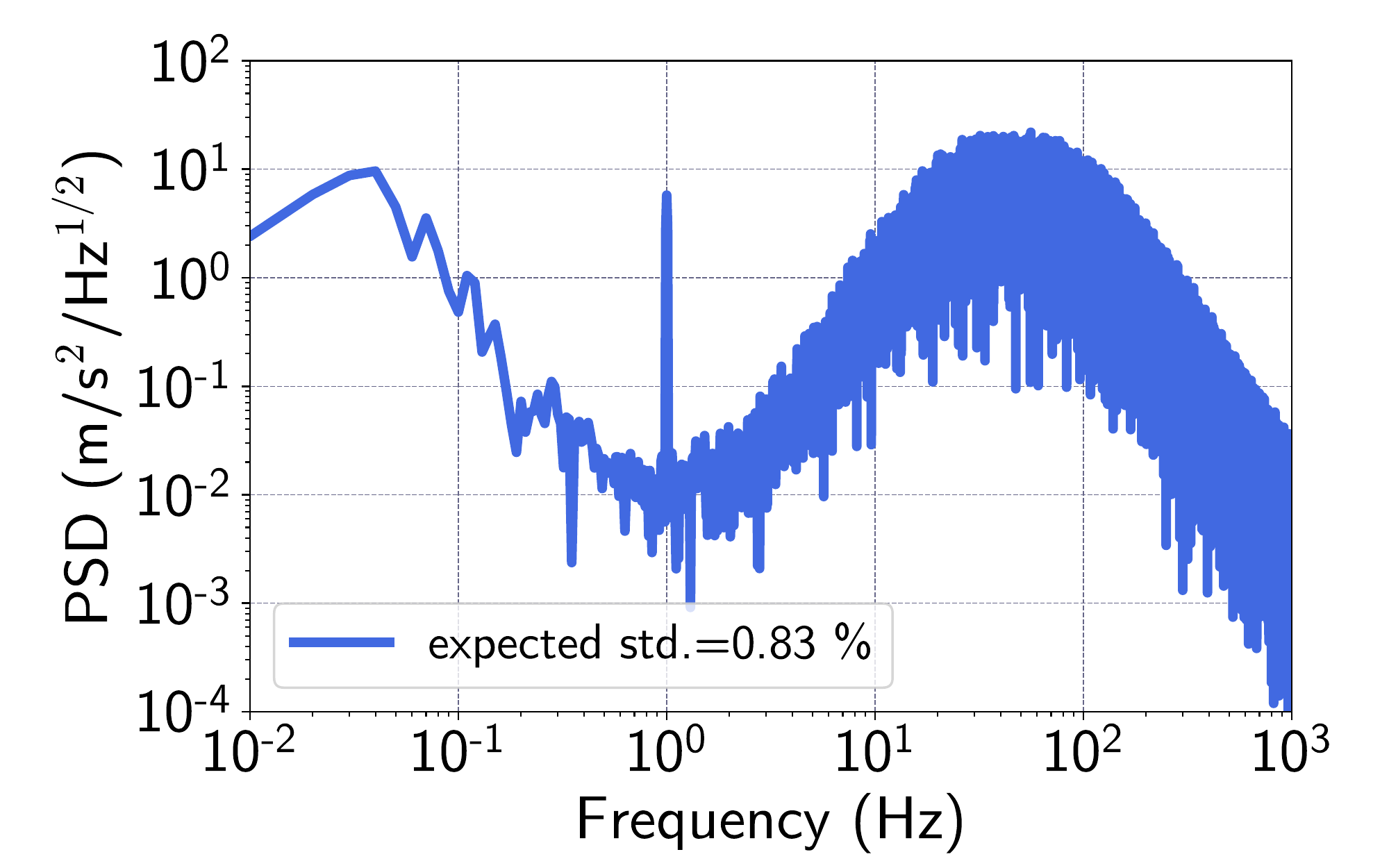}
		\end{center}
	\end{minipage}
	\\
	\begin{minipage}{0.5\hsize}
		\begin{center}
		\includegraphics[width=8cm]{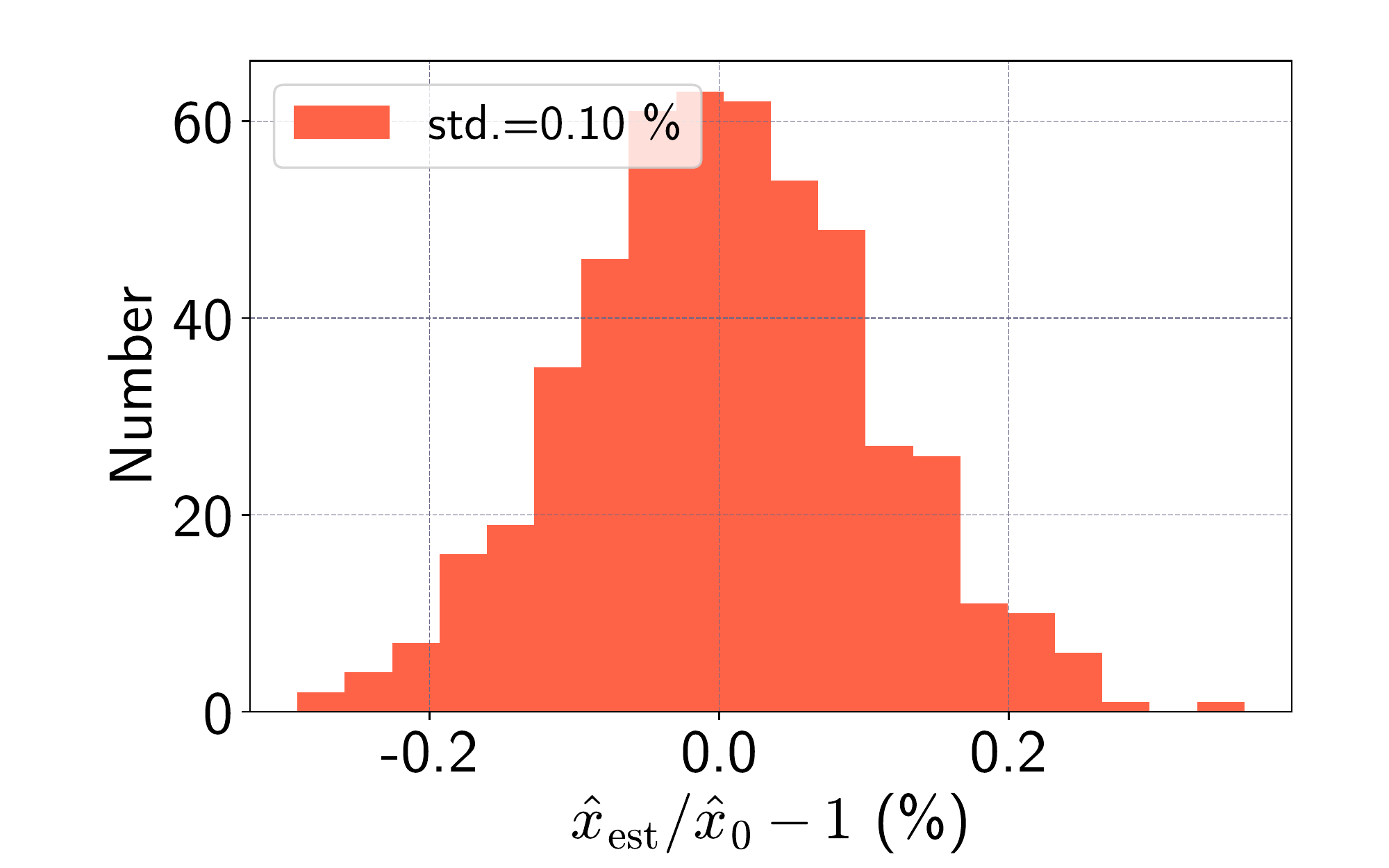}
		\end{center}
	\end{minipage}
	\begin{minipage}{0.5\hsize}
		\begin{center}
		\includegraphics[width=8cm]{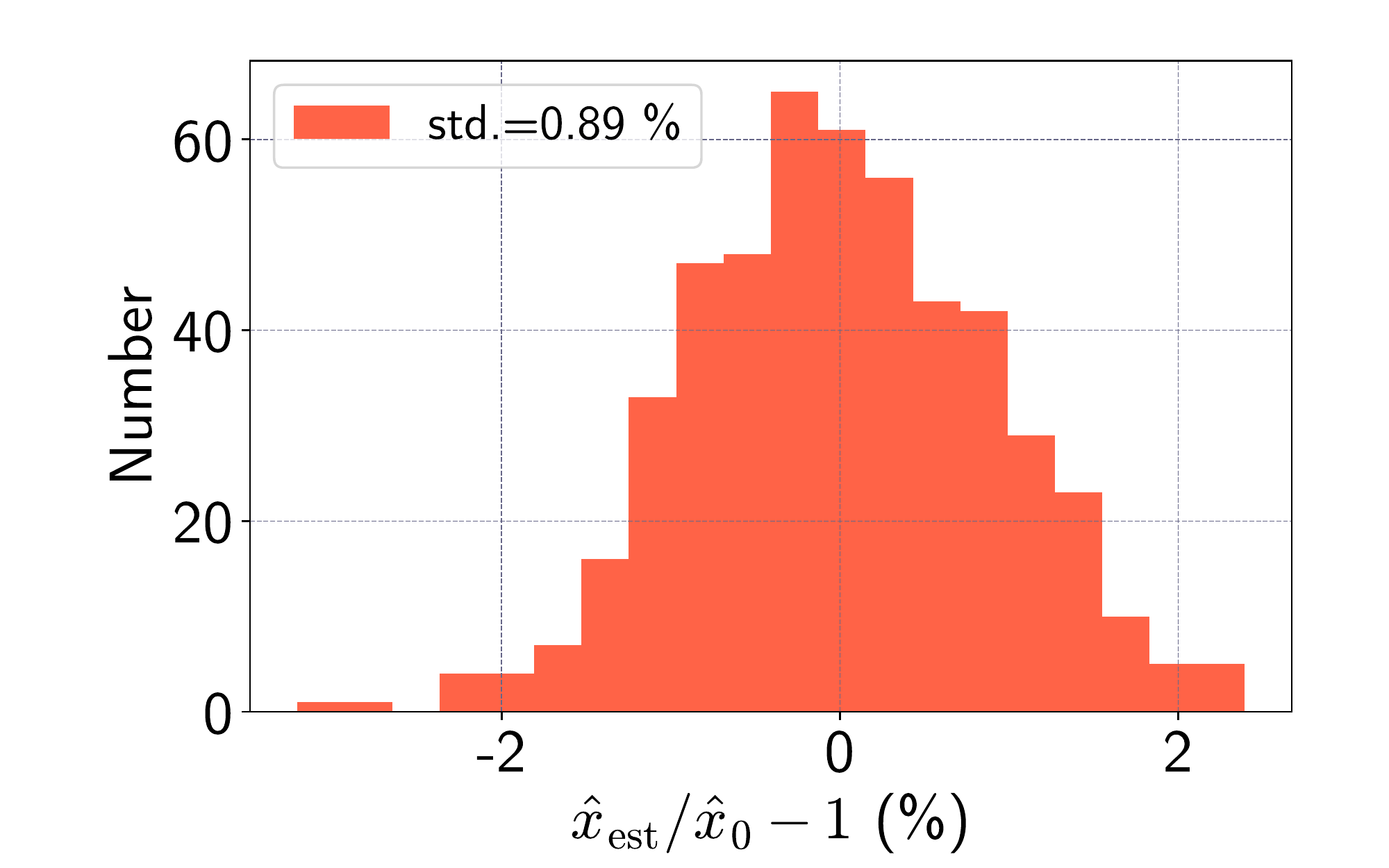}
		\end{center}
	\end{minipage}
\caption{Simulation results of the standard deviation of amplitude estimation for two different noise spectrum cases; flat noise (left column) and frequency-dependent noise (right column). The simulated waveforms (top row), their PSDs (center row), and histograms of the estimated amplitudes (bottom row). The standard deviations expected from Eq.~(\ref{eq:std.n}) are shown in the legend of the PSD figures, and the simulated values are shown in the histogram figures.}
\label{fig:random.simulation}
\end{figure}
To validate the calculations above, a simulation of amplitude estimation was performed.
The excited waveform was $x_0=\hat{x}_0\sin(2\pi f_\mathrm{v}t)$ with $(2\pi f_\mathrm{v})^2\hat{x}_0=1$~m/s$^2$ and $f_\mathrm{v}=1$~Hz.
The data length was set to $T=100$~s.
The noise $n_\mathrm{r}$ was randomly generated 300 times, and the vibration amplitude was estimated from each $x_0 + n_\mathrm{r}$ using Eq.~(\ref{eq:b1b2}).
The standard deviation of the estimated amplitudes $\hat{x}_\mathrm{est}$ relative to the true amplitude $\hat{x}_0$ was calculated and compared to the theoretical expectation from Eq.~(\ref{eq:std.n}).
For the PSD of the noise, $G_\mathrm{r}(f)$, two cases were considered as examples: the flat spectrum $(2\pi f)^2\sqrt{G_\mathrm{r}(f)}\simeq10^{-2}$~(m/s$^2$)/Hz$^{1/2}$ and the frequency-dependent spectrum.
The latter had the same noise levels as the former at the vibration frequency, $(2\pi f_\mathrm{v})^2\sqrt{G_\mathrm{r}(f_\mathrm{v})}\simeq10^{-2}$~(m/s$^2$)/Hz$^{1/2}$, while having a larger noise in $f\neq f_\mathrm{v}$.
Fig.~\ref{fig:random.simulation} shows the simulation results, which agree with the calculation obtained using Eq.~(\ref{eq:std.n}).
The result demonstrates the importance of broadband noise suppression.

\subsubsection{Common random noise for the sensor and reference\\} \label{sec:common.random}
The vibration noise of the exciter $n_x(t)$ is commonly measured by the sensor and reference.
Ideally, such a common noise appears in the two signals in the same way; hence, it does not affect the calibration sensitivity in Eq.~(\ref{eq:Scal}).
However, since the sensor measures $\ddot{n}_x(t)$ while the reference measures $n_x(t)$, the spectral leakage has different contributions for the two signals, which results in $\hat{\ddot{n}}_{x\mathrm{,est}} \neq (2\pi f_\mathrm{v})^2 \hat{n}_{x\mathrm{,est}}$.
From the similar calculations as Section~\ref{sec:indep.random}, the relative standard uncertainty of the sensitivity calibration is given by
\begin{eqnarray}
	\hspace{-10mm} 
	\frac{u(S_\mathrm{cal})}{S} 
	&=& \frac{ u( | \hat{\ddot{n}}_{x\mathrm{,est}} - (2\pi f_\mathrm{v})^2 \hat{n}_{x\mathrm{,est}} | ) }{ (2\pi f_\mathrm{v})^2\hat{x}_0 } \label{eq:uScal.common.0}\\
	&=& \frac{1}{(2\pi f_\mathrm{v})^2\hat{x}_0} \sqrt{ \int_{-\infty}^\infty \frac{|\tilde{W}(f_\mathrm{v}-f)|^2}{T^2} \left\{ (2\pi f)^2 - (2\pi f_\mathrm{v})^2  \right\}^2 G_x(|f|) df }		\label{eq:uScal.common}
\end{eqnarray}
in the absence of the other noise sources.
Here, $G_x(f)$ is the PSD of the common vibration noise $n_x(t)$.
The difference from $n_\mathrm{s}$ and $n_\mathrm{r}$ is that the estimation errors of $\hat{V}_\mathrm{s,est}$ and $\hat{x}_\mathrm{r,est}$ are correlated. 
Eq.~(\ref{eq:uScal.common}) indicates that the acceleration signal amplitude estimation is affected more by the leakage from high-frequency noise than the reference displacement signal.
The same effect can matter when using random or triangle waveform excitation to simultaneously calibrate sensitivity at multiple frequencies.

\begin{figure}
	\begin{minipage}{0.5\hsize}
		\begin{center}
		\includegraphics[width=8cm]{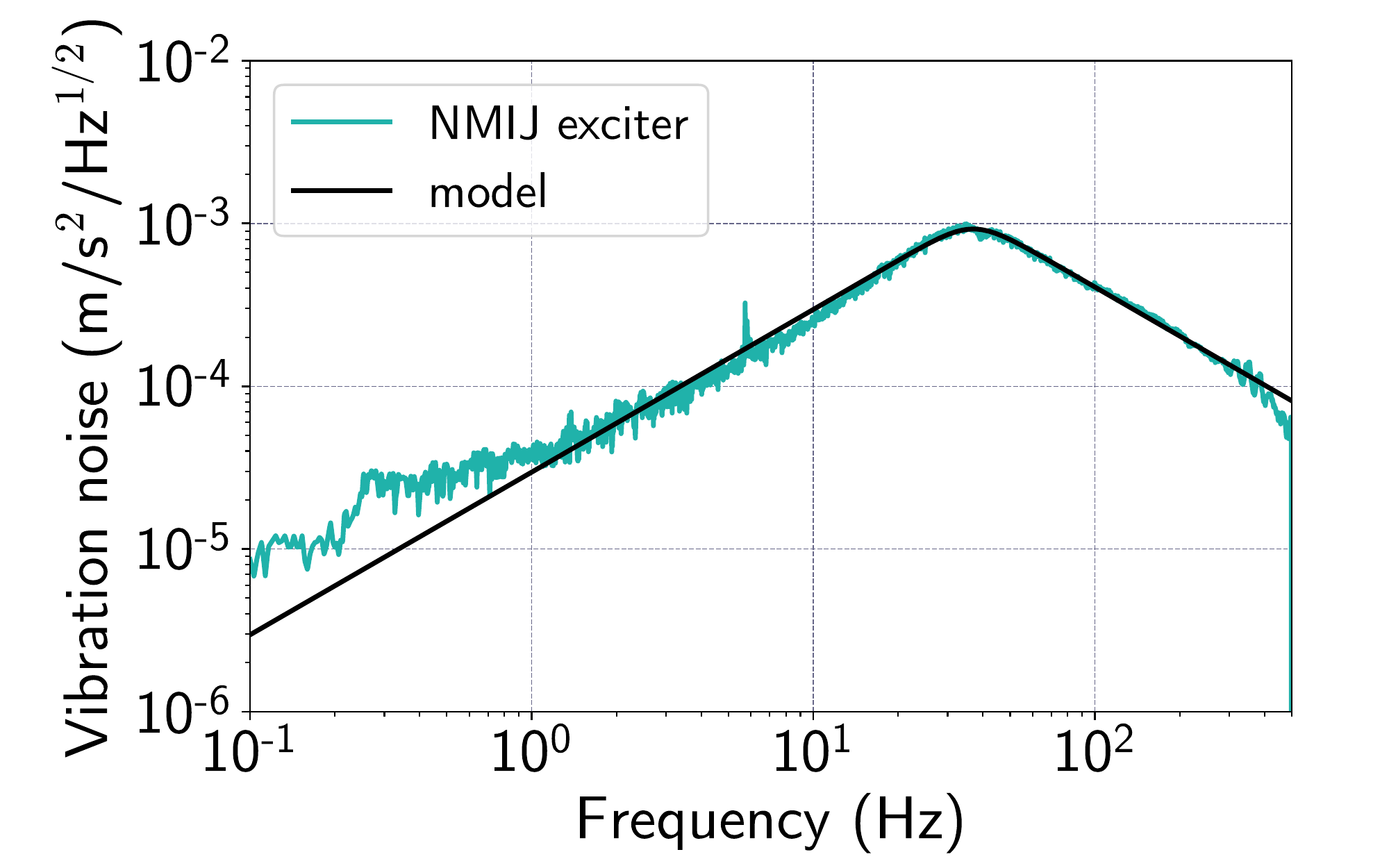}
		\end{center}
	\end{minipage}
	\begin{minipage}{0.5\hsize}
		\begin{center}
		\includegraphics[width=8cm]{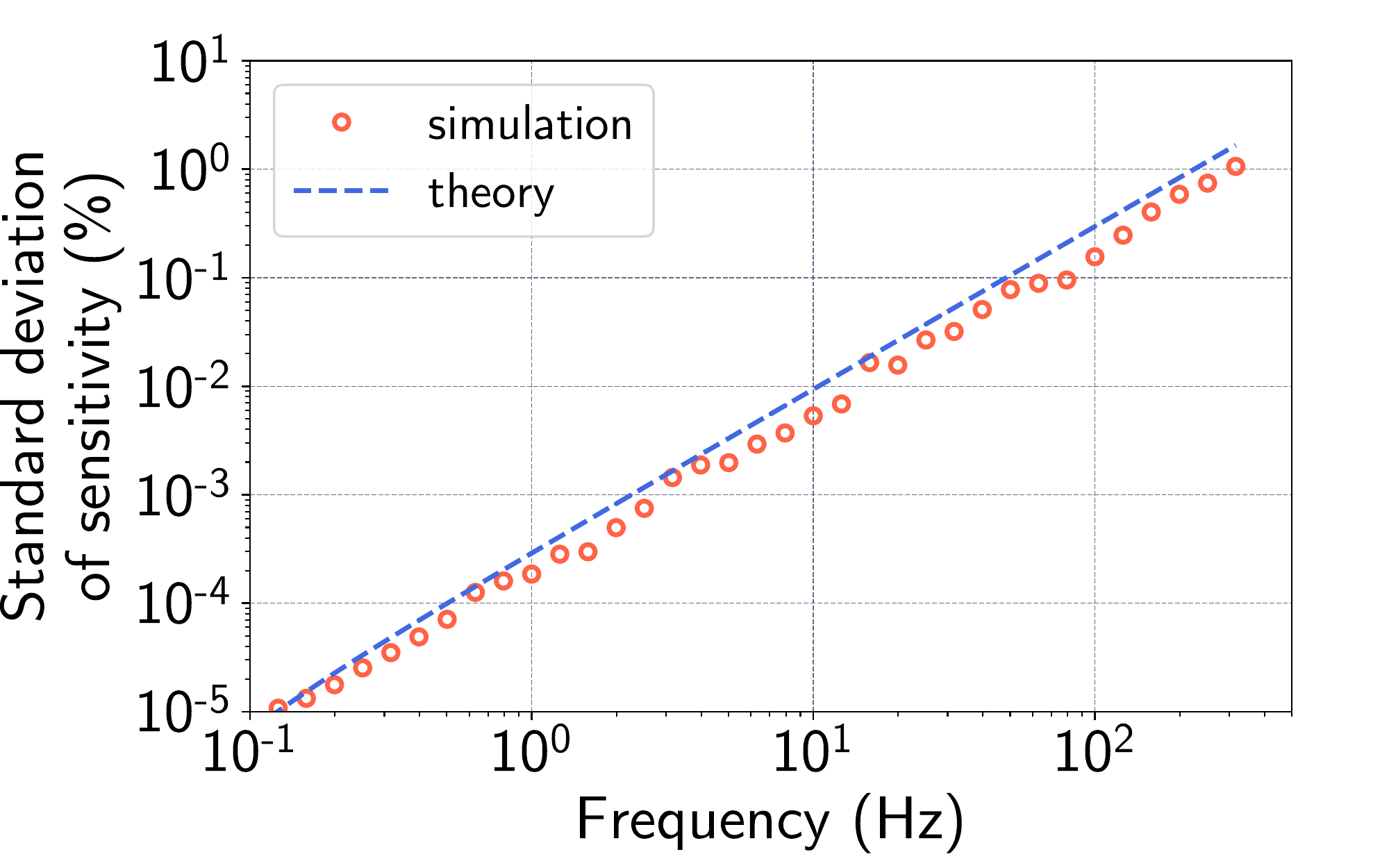}
		\end{center}
	\end{minipage}
\caption{(left) Amplitude spectral density of the vibration noise in the low-frequency vibration exciter of NMIJ. The measured spectrum (green) and model (black) used for the calculations. (right) Standard deviation of the estimated sensitivity under the random vibration noise. The results of simulation (red circles) and calculation using Eq.~(\ref{eq:uScal.common}) (blue dashed-line).}
\label{fig:common.simulation}
\end{figure}
To validate the calculation and show the order of the uncertainty contribution, we performed a simulation similar to that shown in Section~\ref{sec:indep.random}.
The excited acceleration amplitude was fixed to $(2\pi f_\mathrm{v})^2\hat{x}_0=1$~m/s$^2$ for frequency varying from 0.1~Hz to 300~Hz.
The data length was set to $T=100/f_\mathrm{v}$.
The background vibration spectrum in the acceleration unit, $(2\pi f)^2\sqrt{G_x(f)}$, of the calibration system in NMIJ is shown in Fig.~\ref{fig:common.simulation}.
The smoothed spectrum model was used for the simulation. 
The common vibration noise $n_x$ was randomly generated 20 times at each frequency, and the amplitude of the reference displacement, $\hat{x}_\mathrm{r,est}$, was estimated from each $x_0 + n_\mathrm{r}$ using Eq.~(\ref{eq:b1b2}).
Then, the same time series were numerically differentiated twice to prepare the sensor signal $S(\ddot{x}_0 + \ddot{n}_\mathrm{r})$, which was used to estimate the amplitude $\hat{V}_\mathrm{s,est}$.
The standard deviation of the estimated amplitude ratio $\hat{V}_\mathrm{s,est}/(2\pi f_\mathrm{v})^2\hat{x}_\mathrm{r,est}$ relative to the true sensitivity modulus $S$ was calculated at each frequency and compared with the theoretical result obtained from Eq.~(\ref{eq:uScal.common}).
The results are shown in Fig.~\ref{fig:common.simulation}. 
Eq.~(\ref{eq:uScal.common}) explains the simulated standard deviation of the sensitivity modulus.
The contribution ranged from 0.1~\% to 1~\% around 100~Hz, which is not ignorable in accelerometer calibration.

\subsection{Effect of line noise and harmonics} \label{sec:SAM.line}
The error components include both the line noise and the harmonics of the input vibration having constant amplitudes and phases.
The difference between them is that the frequency of the former is fixed and independent of $f_\mathrm{v}$, while the latter always appears at integer multiples of $f_\mathrm{v}$.
Their contributions can be treated in the same way using Eq.~(\ref{eq:c}).
Here, we consider the general case of the line noise $l(t)=\hat{l} \sin(2\pi f_l t+\phi_l)$ added to a sinusoidal wave $x_0(t)=\hat{x}_0 \sin(2\pi f_\mathrm{v}t)$.
The complex amplitude estimated from $x(t)=x_0(t)+l(t)$ is calculated from Eq.~(\ref{eq:c}) as 
\begin{equation}
	\hat{x}_\mathrm{est} = i \hat{x}_0 + i \hat{l} \left( \frac{\tilde{W}(f_\mathrm{v}-f_l)}{T} e^{-i\phi_l} - \frac{\tilde{W}(f_\mathrm{v}+f_l)}{T} e^{i\phi_l} \right). \label{eq:xest.line}
\end{equation}
Then, the amplitude and phase estimation errors are
\begin{eqnarray}
	\frac{|\hat{x}_\mathrm{est}|}{\hat{x}_0}-1 &=& \frac{\hat{l}}{\hat{x}_0} \frac{\tilde{W}(f_\mathrm{v}-f_l)-\tilde{W}(f_\mathrm{v}+f_l)}{T} \cos(\phi_l), \\
	\arg[\hat{x}_\mathrm{est}]  - \frac{\pi}{2} &=& \frac{\hat{l}}{\hat{x}_0} \frac{\tilde{W}(f_\mathrm{v}-f_l)+\tilde{W}(f_\mathrm{v}+f_l)}{T} \sin(\phi_l).
\end{eqnarray}
In actual measurements, the phase $\phi_l$ of the line noise relative to $x_0$ is random for each measurement and is uniformly distributes from 0 to $2\pi$.
Therefore, their standard uncertainties are given by
\begin{eqnarray}
	\frac{u(|\hat{x}_\mathrm{est}|)}{\hat{x}_0} &=& \frac{\hat{l}}{\sqrt{2}\hat{x}_0} \frac{|\tilde{W}(f_\mathrm{v}-f_l)-\tilde{W}(f_\mathrm{v}+f_l)|}{T}, \label{eq:ux.line.amplitude}	\\
	u(\arg[\hat{x}_\mathrm{est}]) &=& \frac{\hat{l}}{\sqrt{2}\hat{x}_0} \frac{|\tilde{W}(f_\mathrm{v}-f_l)+\tilde{W}(f_\mathrm{v}+f_l)|}{T}. \label{eq:ux.line.phase}
\end{eqnarray}
The uncertainty follows a U-shaped distribution.
Since $\tilde{W}(f)$ has large value around $f\simeq0$, the uncertainty is large when $f_\mathrm{v}\simeq f_l$.
In this case, Eqs.~(\ref{eq:ux.line.amplitude}) and (\ref{eq:ux.line.phase}) are approximated as 
\begin{equation}
	\frac{u(|\hat{x}_\mathrm{est}|)}{\hat{x}_0} \simeq u(\arg[\hat{x}_\mathrm{est}]) \simeq \frac{\hat{l}}{\sqrt{2}\hat{x}_0} \frac{|\tilde{W}(f_\mathrm{v}-f_l)|}{T}. \label{eq:ux.line.approx}
\end{equation}

Notably, the harmonics of $f_\mathrm{v}$ do not affect the amplitude estimation through the spectral leakage when the rectangular window is used because $\tilde{W}_\mathrm{r}(f_\mathrm{v}\pm Nf_\mathrm{v})=0$ ($N$: integer) if the length $T$ is integer multiples of the vibration period.
It is the same for the Hanning window or some other types of windows.
Note that it does not mean that the harmonics do not become an error source in any case.
If the sensor or reference interferometer has nonlinearity, the harmonics $l_x$ is nonlinearly converted with the fundamental wave $x_0$ to the line noise $l_\mathrm{s}$ or $l_\mathrm{r}$ at $f_\mathrm{v}$ and can affect the estimated amplitude even if $T$ is properly selected.
To evaluate such a contribution, detailed information about the input harmonics $l_x$ and the nonlinearity of the sensor are necessary, which is out of the scope of this work.
As far as the harmonics $l_x$, $l_\mathrm{s}$, and $l_\mathrm{r}$ are independent from each other, the contribution of the harmonics can be easily eliminated by the proper choice of $T$.
Therefore, we mainly discuss the line noise in the remainder of this article.

\subsubsection{Independent line noise for the sensor and reference\\} \label{sec:indep.line}
The standard uncertainties of sensitivity under the independent line noise of the sensor and reference, $l_\mathrm{s}=\hat{l}_\mathrm{s}\sin(2\pi f_{l, \mathrm{s}}t + \phi_{l, \mathrm{s}})$ and $l_\mathrm{r}=\hat{l}_\mathrm{r}\sin(2\pi f_{l, \mathrm{r}}t + \phi_{l, \mathrm{r}})$, are given by the square-root of the square sum of their uncertainty contributions.
Using Eq.~(\ref{eq:ux.line.approx}), it is approximated as 
\begin{equation}
	\hspace{-10mm}
	\frac{u(S_\mathrm{cal})}{S} 
	= \frac{1}{\sqrt{2}} \sqrt{ \left( \frac{\hat{l}_\mathrm{s}}{(2\pi f_\mathrm{v})^2 \hat{x}_0} \right)^2 \frac{|\tilde{W}(f_\mathrm{v}-f_{l, \mathrm{s}})|^2}{T^2} + \left( \frac{\hat{l}_\mathrm{r}}{\hat{x}_0} \right)^2 \frac{|\tilde{W}(f_\mathrm{v}-f_{l, \mathrm{r}})|^2}{T^2} }. \label{eq:uScal.line.indep}
\end{equation}
Here, we assumed that $\phi_{l, \mathrm{s}}$ and $\phi_{l, \mathrm{r}}$ are not correlated to each other.
Accurate expression is given by replacing $\tilde{W}(f_\mathrm{v}-f_l)$ with $\tilde{W}(f_\mathrm{v}-f_l)-\tilde{W}(f_\mathrm{v}+f_l)$ for sensitivity and $\tilde{W}(f_\mathrm{v}-f_l)-\tilde{W}(f_\mathrm{v}+f_l)$ for phase delay.

When the line noise frequency is equal to the vibration frequency, $\tilde{W}(f_\mathrm{v}-f_l)=T$.
Therefore, the relative amplitude estimation uncertainties of the sensor and reference are determined by only the amplitude ratio:
\begin{equation}
	\frac{u(|\hat{x}_\mathrm{r, est}|)}{\hat{x}_0} = \frac{\hat{l}_\mathrm{r}}{\sqrt{2}\hat{x}_0}.
\end{equation}
This is the fundamental limit of uncertainty from the line noise.
Although the leakage from $f\neq f_\mathrm{v}$ can be reduced by proper signal processing, the contribution at $f=f_\mathrm{v}$ cannot be avoided unless the line noise amplitude is reduced.

\begin{figure}
	\begin{minipage}{0.5\hsize}
		\begin{center}
		\includegraphics[width=8cm]{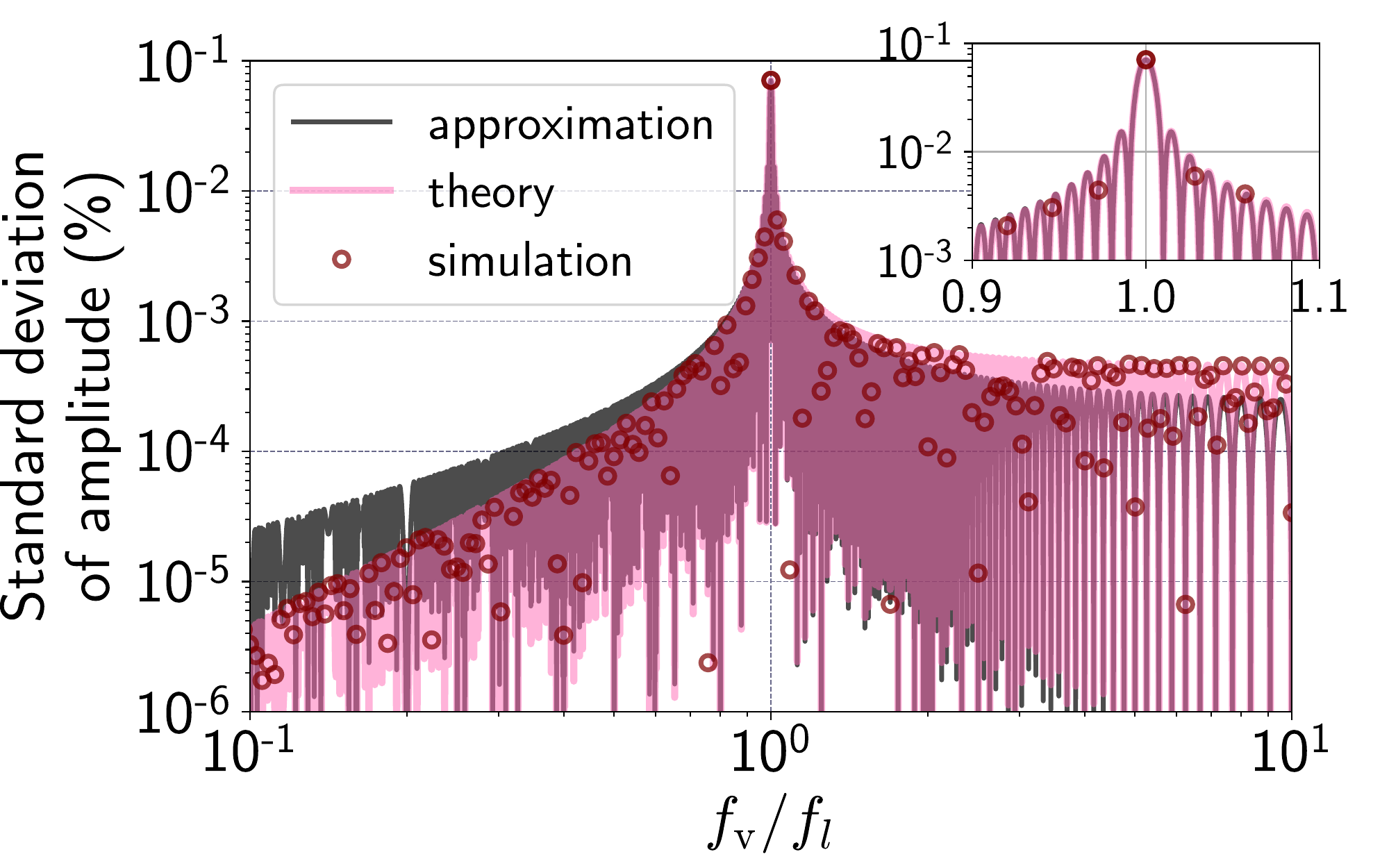}
		\end{center}
	\end{minipage}
	\begin{minipage}{0.5\hsize}
		\begin{center}
		\includegraphics[width=8cm]{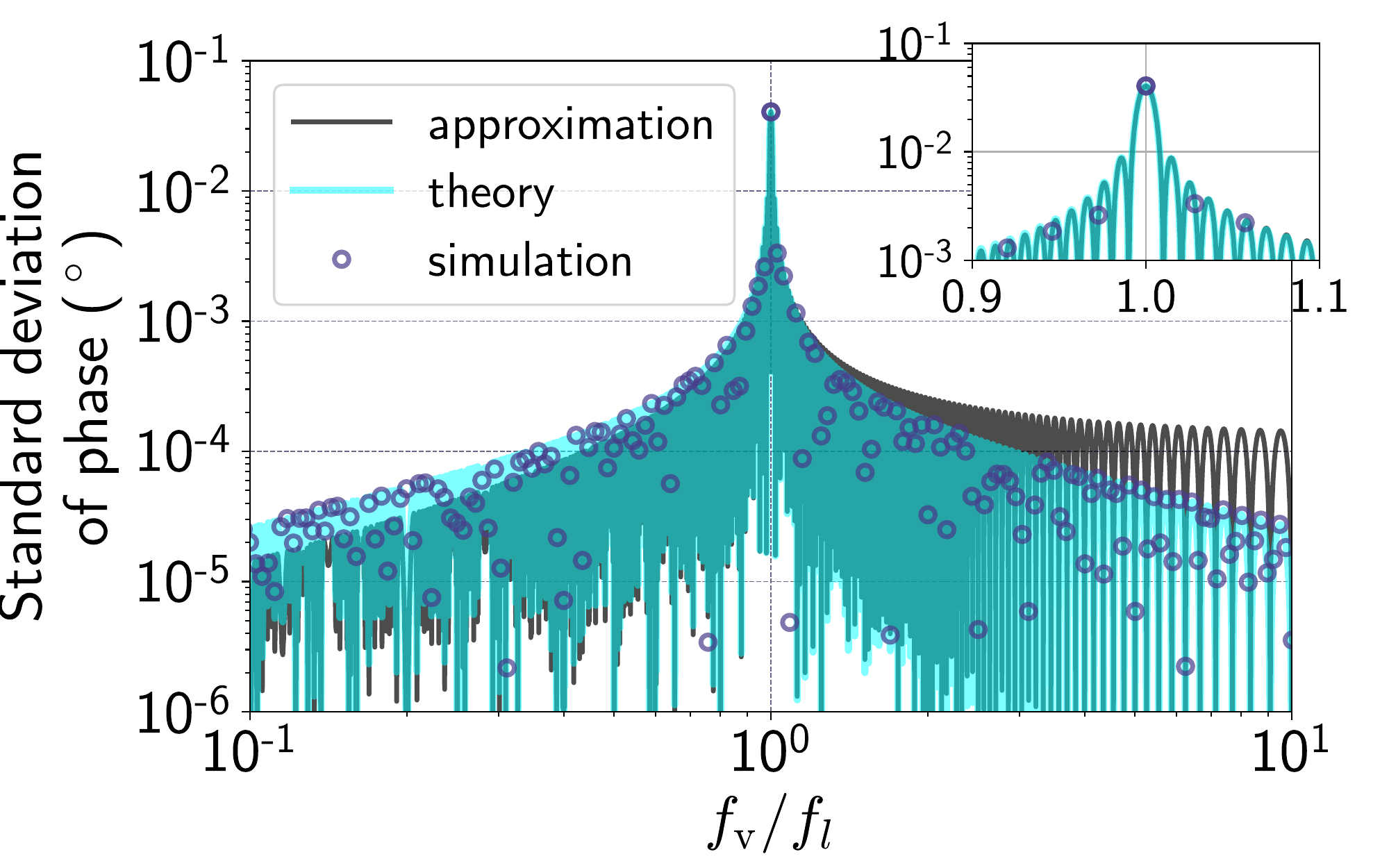}
		\end{center}
	\end{minipage}
	\caption{Relative amplitude (left) and phase (right) estimation uncertainty due to the line noise in the case of $\hat{l}/\hat{x}_0=10^{-3}$. The open circles show the simulated result, the solid magenta and cyan lines show the theoretical uncertainty expected from Eqs.~(\ref{eq:ux.line.amplitude}) and (\ref{eq:ux.line.phase}), respectively, and the black lines show the approximated expression Eq.~(\ref{eq:ux.line.approx}).}
	\label{fig:linenoise.50Hz}
\end{figure}
The line noise contribution was confirmed by simulation.
Fixing $\hat{l}/\hat{x}_0=10^{-3}$, we estimated the amplitude and phase of $x(t)=x_0(t)+l(t)$ for various $f_\mathrm{v}/f_l$ and $\phi_l$.
The data length was fixed to $T=100/f_\mathrm{v}$ at each frequency.
The standard deviations for different $\phi_l$ were calculated from the simulated values and compared with Eqs.~(\ref{eq:ux.line.amplitude}) and (\ref{eq:ux.line.phase}).
Fig.~\ref{fig:linenoise.50Hz} shows the simulation results.
Theoretical calculations explained the uncertainty well.
The approximated formula, Eq.~(\ref{eq:ux.line.approx}), also accurately estimated the uncertainty around the line noise frequency $f_l$, where the line noise contribution becomes important.
At frequencies away from $f_l$, Eq.~(\ref{eq:ux.line.approx}) led to over-/under-estimation by a few times.

\subsubsection{Common line noise for the sensor and reference\\} \label{sec:common.line}
The standard uncertainty of the sensitivity under the common line noise $l_x=\hat{l}_x \sin(2\pi f_{l, x}t + \phi_{l, x})$ was calculated similarly as Eq.~(\ref{eq:uScal.common.0}).
Using Eq.~(\ref{eq:ux.line.approx}), it was approximated as
\begin{equation}
	\hspace{-10mm}
	\frac{u(S_\mathrm{cal})}{S} \simeq u(\Delta\phi_\mathrm{cal})
	\simeq \frac{\hat{l}_x}{\sqrt{2}\hat{x}_0} \frac{|f_\mathrm{v}^2-f_{l,x}^2|}{f_\mathrm{v}^2} \frac{|\tilde{W}(f_\mathrm{v}-f_{l,x})|}{T}. \label{eq:uScal.line.common}
\end{equation}
Again, accurate expressions are given by replacing $\tilde{W}(f_\mathrm{v}-f_{l,x})$ with $\tilde{W}(f_\mathrm{v}-f_{l,x})\pm\tilde{W}(f_\mathrm{v}+f_{l,x})$.

\begin{figure}
	\begin{minipage}{0.5\hsize}
		\begin{center}
		\includegraphics[width=8cm]{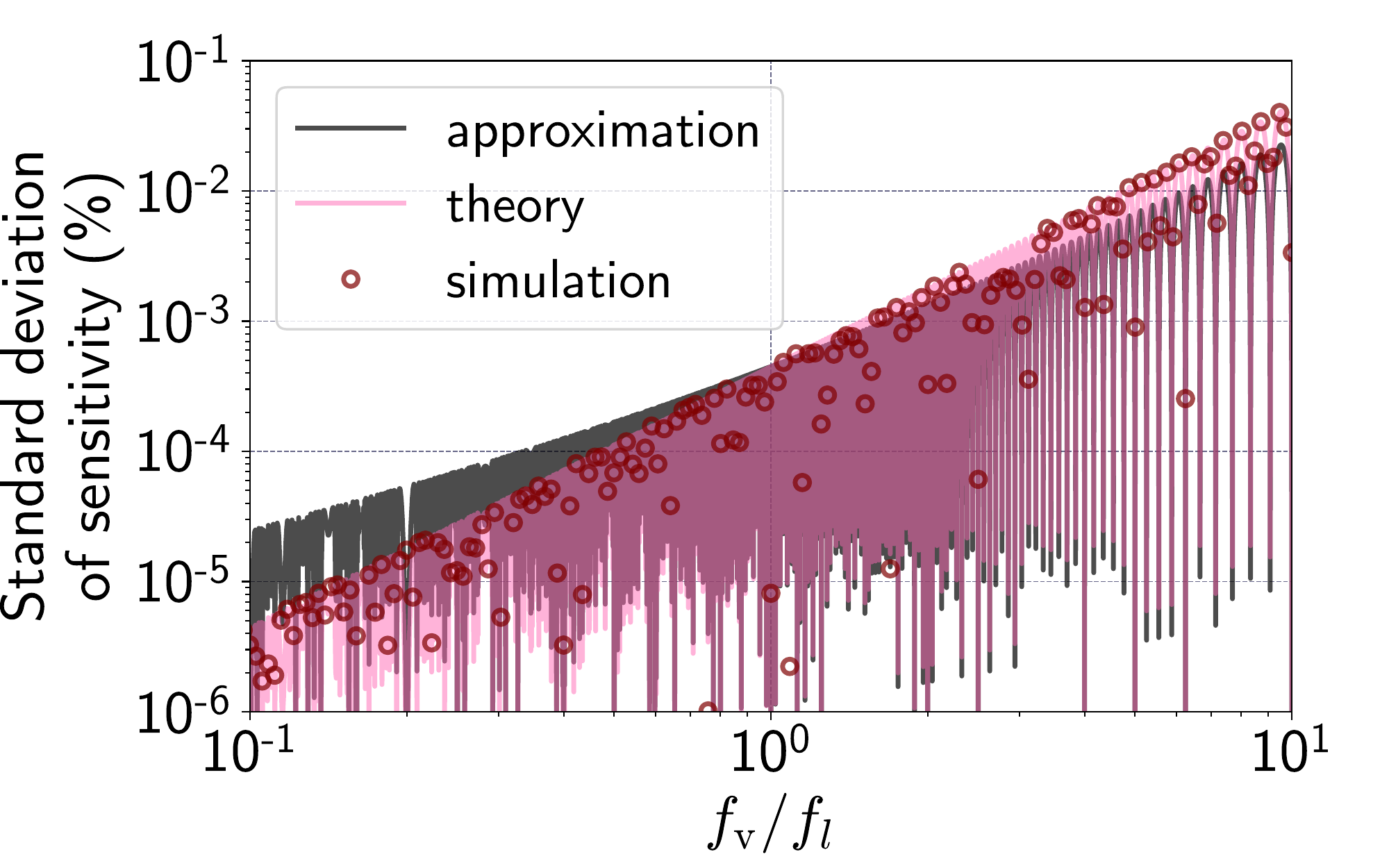}
		\end{center}
	\end{minipage}
	\begin{minipage}{0.5\hsize}
		\begin{center}
		\includegraphics[width=8cm]{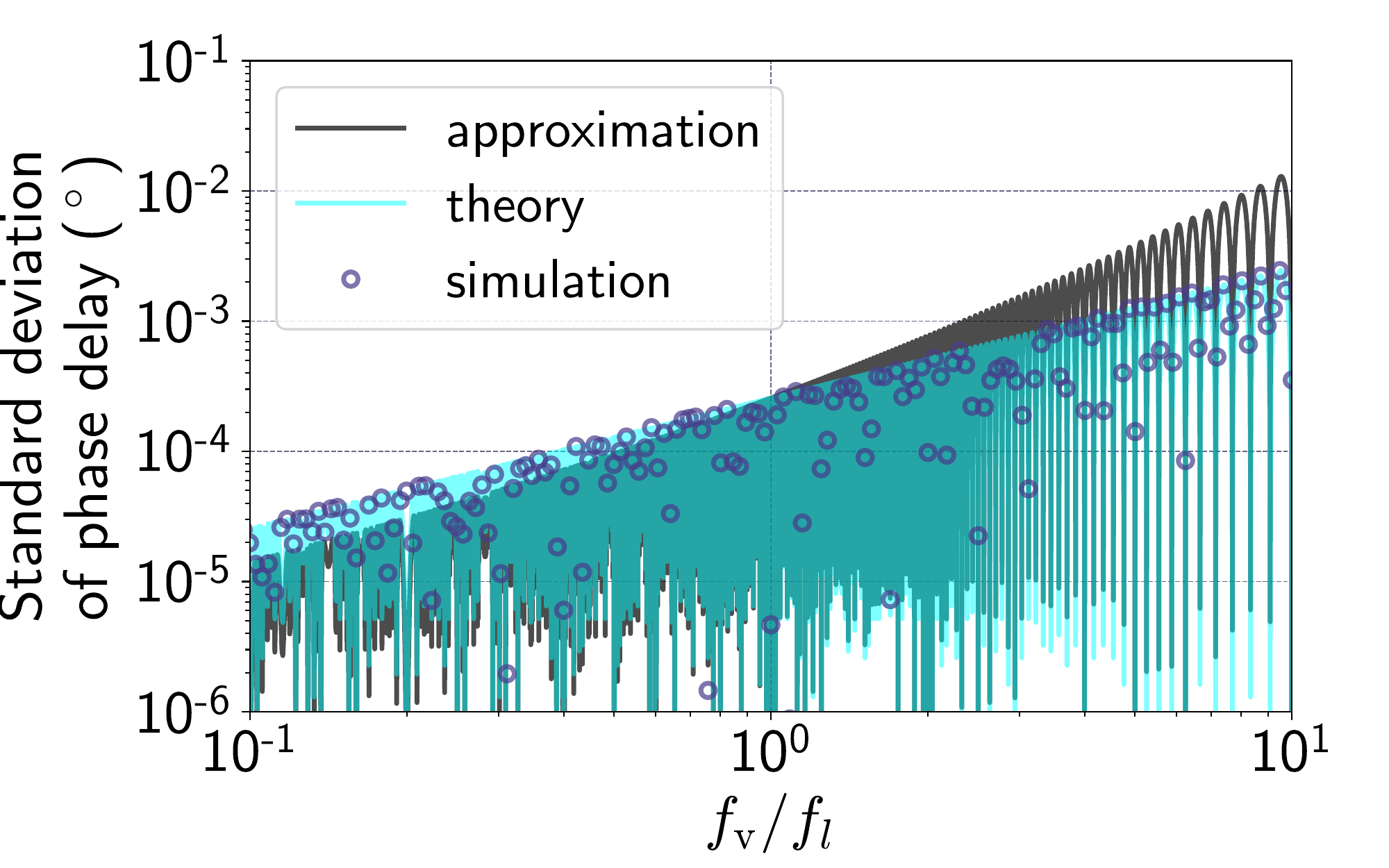}
		\end{center}
	\end{minipage}
	\caption{Sensitivity (left) and phase delay (right) estimation uncertainty due to the common line noise $l_x$ with $\hat{\ddot{l}}_x/\hat{\ddot{x}}_0=10^{-3}$. The approximation formula Eq.~(\ref{eq:uScal.line.common}) is shown with a black line, and the accurate formulae are shown with red and blue lines. The simulated values are plotted with open circles.}
	\label{fig:linenoise.common.50Hz}
\end{figure}
The effect of the line noise was simulated again.
The common line noise amplitude was fixed as $\hat{\ddot{l}}_x/\hat{\ddot{x}}_0=10^{-3}$, and the data length was fixed to $T=100/f_\mathrm{v}$ at each frequency.
The sensor and reference signals $V_\mathrm{s}=S(\ddot{x}_0+\ddot{l}_x)$ and $x_\mathrm{r}=x_0+l_x$ were prepared for different line phase $\phi_{l,x}$. 
The standard deviations of the estimated sensitivity and phase for different $\phi_{l,x}$ were calculated.
The simulation results are shown in Fig.~\ref{fig:linenoise.common.50Hz}.
The results agreed with the theoretical calculations using Eq.~(\ref{eq:uScal.line.common}).
Unlike the independent line noise, the uncertainty contribution of the common line noise is not concentrated around $f_\mathrm{v}\simeq f_{l,x}$, because the peak of $\tilde{W}(f_\mathrm{v}-f_{l,x})$ is canceled by the factor $f_\mathrm{v}^2-f_{l,x}^2$, as shown in Eq.~(\ref{eq:uScal.line.common}), when the rectangular window is used.

\section{Reduction of noise contribution in calibration}	\label{sec:reduce}
The calibration uncertainty considered in this paper is explained by Eqs.~(\ref{eq:uScal.random}), (\ref{eq:uScal.common}), (\ref{eq:uScal.line.indep}), and (\ref{eq:uScal.line.common}).
These equations show how the spectral leakage contributes to the calibration in the conventional acceleration calibration with the SAM.
As already mentioned, the background random/line noise of the sensing parts (sensor and interferometer) at $f_\mathrm{v}$ is a fundamental limit of signal processing.
In this section, we aimed to minimize the leakage from $f\neq f_\mathrm{v}$.
In the following subsections, we propose three signal processing modification methods including filtering, changing the window function, and numerical differentiation to align the unit of the measurand.
For the line noise, the selection of the data length $T$ is also discussed.
The investigations about the three modifications are mainly focused on the random noise, although the same methods are applicable to the line noise.
These methods are simulated for both random and line noises.

\subsection{Filtering the signal} \label{sec:filter}
One of the simplest ways to reduce the leakage is filtering the signal before amplitude estimation.
The bandpass filter (BPF) centered at $f_\mathrm{v}$ reduces the noise PSD $G(f\neq f_\mathrm{v})$ in Eqs.~(\ref{eq:uScal.random}) and (\ref{eq:uScal.common}) and the line noise at $f\neq f_\mathrm{v}$.
Depending on the background noise spectrum, the low-pass or high-pass filters can also be used.
Such filtering is already adopted in the calibration process in NMIJ, although the shape of the filter has been empirically determined.
The filtering changes $G(f\neq f_\mathrm{v})$ in Eqs.~(\ref{eq:uScal.random}) and (\ref{eq:uScal.common}) to $|\tilde{F}(f)|^2G(f)$, where $\tilde{F}(f)$ is the transfer function of the filter.
The line noise amplitude also decreases to $|\tilde{F}(f)|\hat{l}$ in Eqs.~(\ref{eq:uScal.line.indep}) and (\ref{eq:uScal.line.common}).

\begin{figure}
	\begin{center}
	\includegraphics[width=10cm]{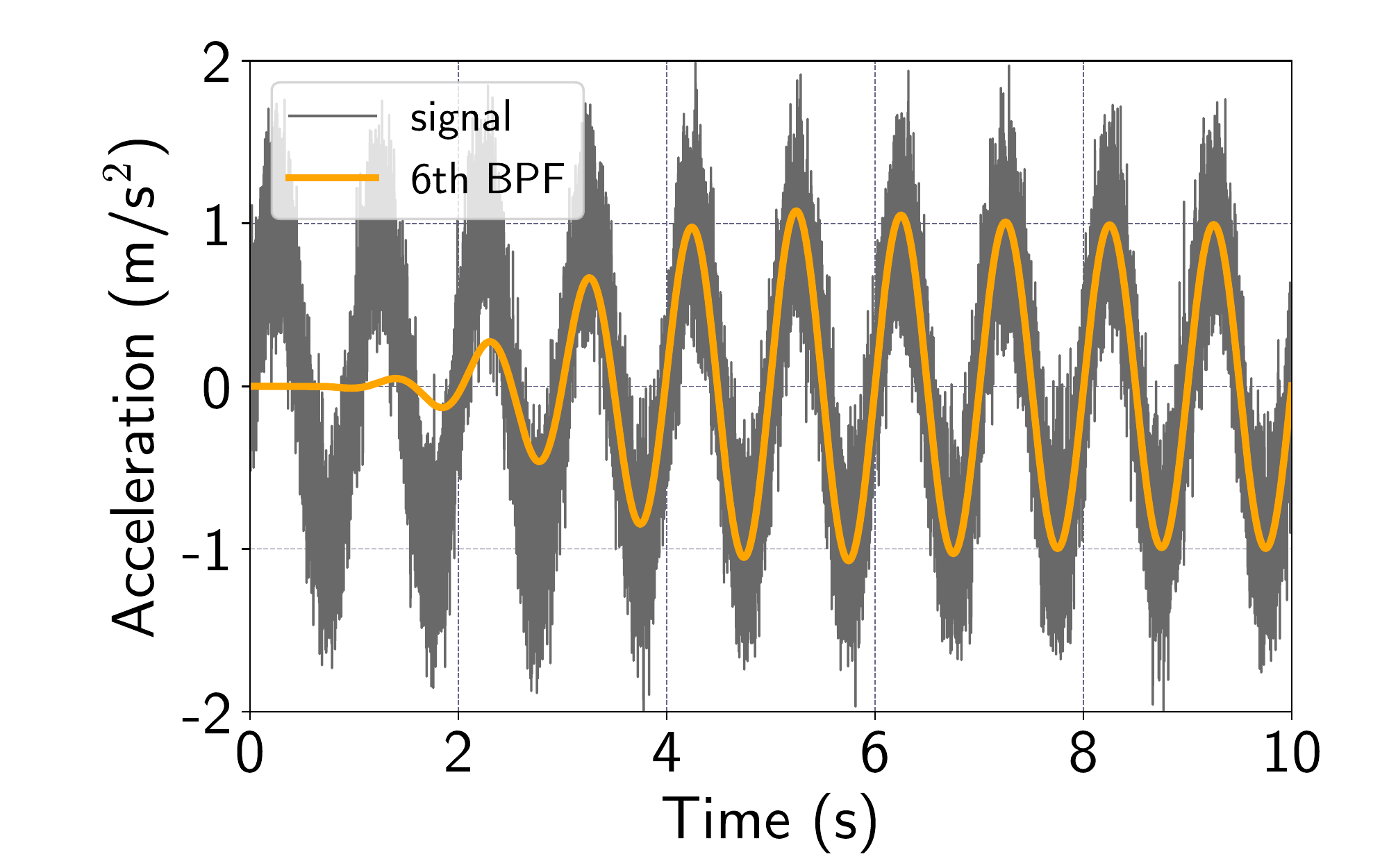}
	\caption{The noisy waveform with the constant sinusoidal amplitude starting at $t=0$ (grey) and bandpass-filtered data (orange).}
	\label{fig:filtered.wave}
	\end{center}
\end{figure}
The disadvantage of filtering is that it takes time until the waveform becomes stable after starting the excitation.
An example is shown in Fig.~\ref{fig:filtered.wave}.
A sixth-order BPF was applied to the noisy waveform to extract the frequency component around $f_\mathrm{v}$ with a Q-factor ($Q$) of $\simeq2$. 
Here, the Q-factor is the ratio of the center frequency $f_\mathrm{v}$ over the bandwidth of the filter.
In the example, the amplitude of the filtered data became stable after about eight vibration cycles.
Consequently, the first several vibration cycles of the data could not be used for amplitude estimation, which increased the measurement time, especially at low frequencies.
\begin{figure}
	\begin{minipage}{0.5\hsize}
		\begin{center}
		\includegraphics[width=8cm]{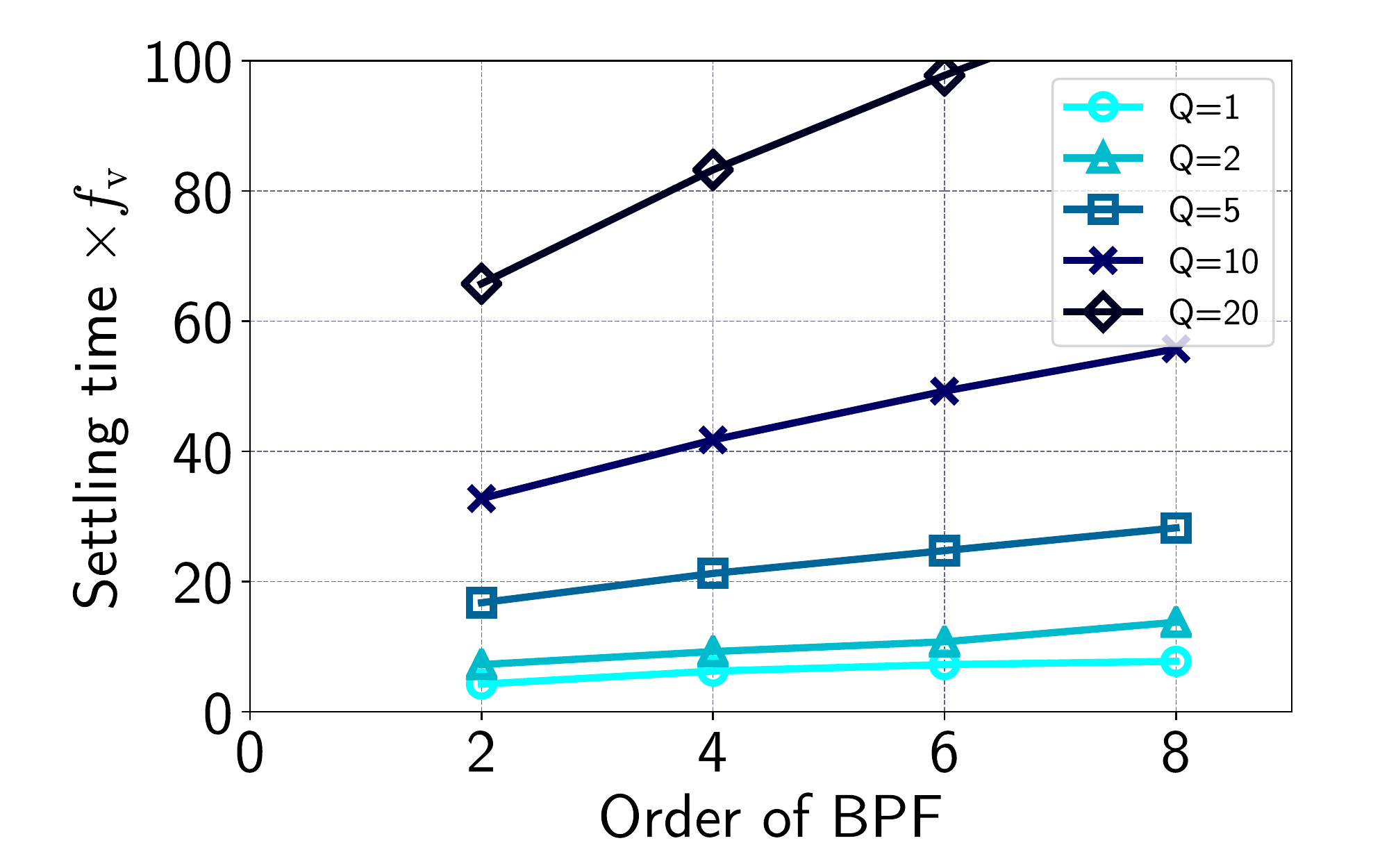}
		\end{center}
	\end{minipage}
	\begin{minipage}{0.5\hsize}
		\begin{center}
		\includegraphics[width=8cm]{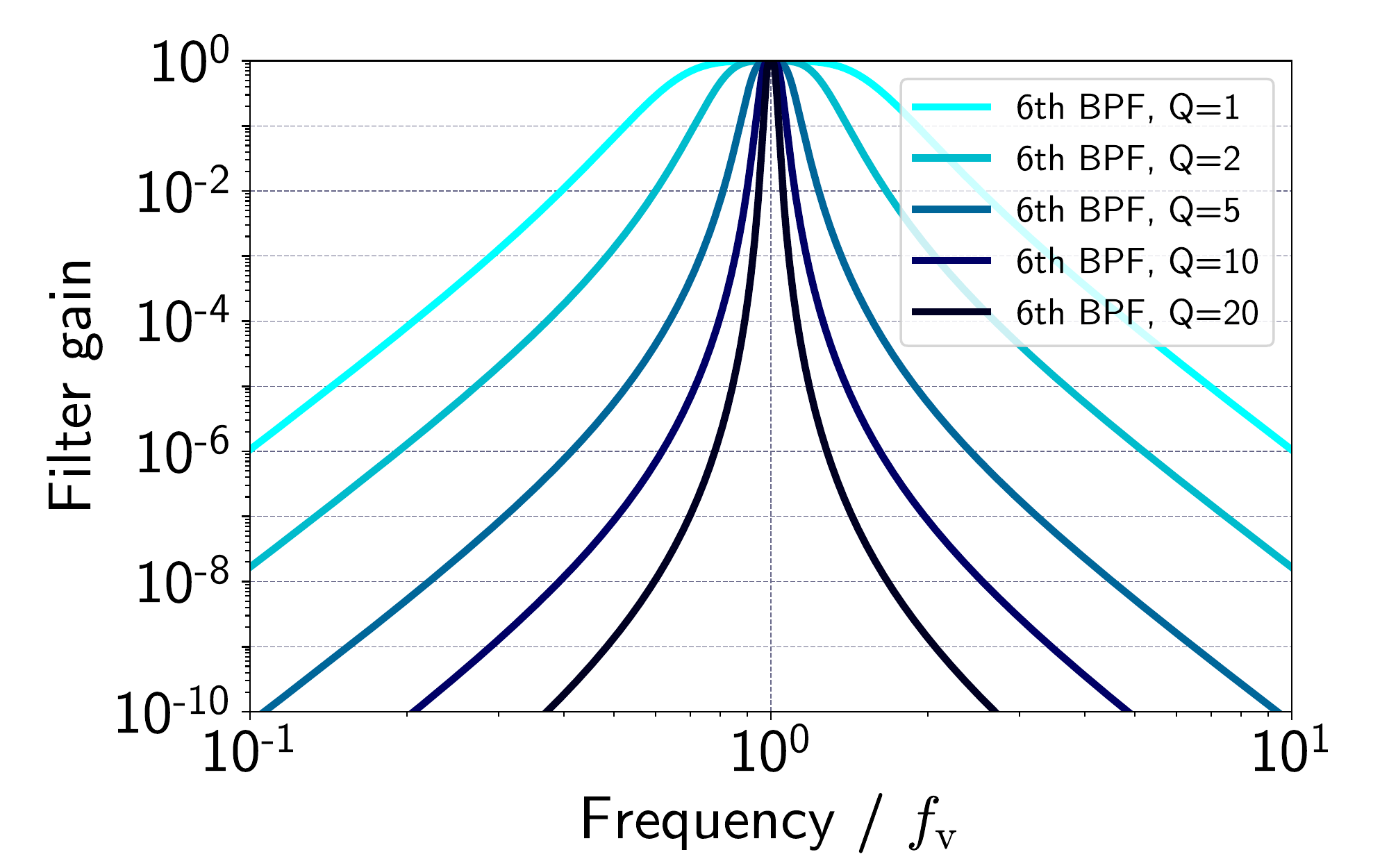}
		\end{center}
	\end{minipage}
\caption{(left) Simulated settling times of bandpass filters with different combinations of the order (2, 4, 6, and 8) and Q-factor (1, 2, 5, 10, and 20). (right) The filter gain of the sixth-order bandpass filters used for the calculation.}
\label{fig:filter.settling}
\end{figure}
Fig.~\ref{fig:filter.settling} shows the simulated settling time for different orders and Q-factors of the BPF.
The settling time is defined as the time length until the filtered waveform amplitude settles within $\pm0.1$~\% of the true amplitude.
The settling time is roughly proportional to the Q-factor, and increases with the order of the filter.
Thus, a trade-off exists between the noise reduction ability of the filter and discarded measurement time.
Although it may be possible to compensate for the amplitude change associated with settling, it may take effort to perform a simulation according to the shape of the filter and the input signal.
In the remainder of this article, we use a sixth-order BPF with $Q=1$ to keep the settling time within 10 vibration cycles.

\subsection{Changing the window function}\label{sec:window}
Another way to improve the vibration amplitude estimation accuracy is changing the window function $w(t)$.
The rectangular window, which is implicitly used in the conventional SAM, is famous for large spectral leakage.
The other window functions such as a Hanning window can reduce $\tilde{W}(f)$ at $f\neq f_\mathrm{v}$.
Although this method is well-known for fast Fourier transform applications, it has not been discussed for the SAM in vibration calibration.

\begin{figure}
	\begin{minipage}{1\hsize}
		\begin{center}
		\includegraphics[width=10cm]{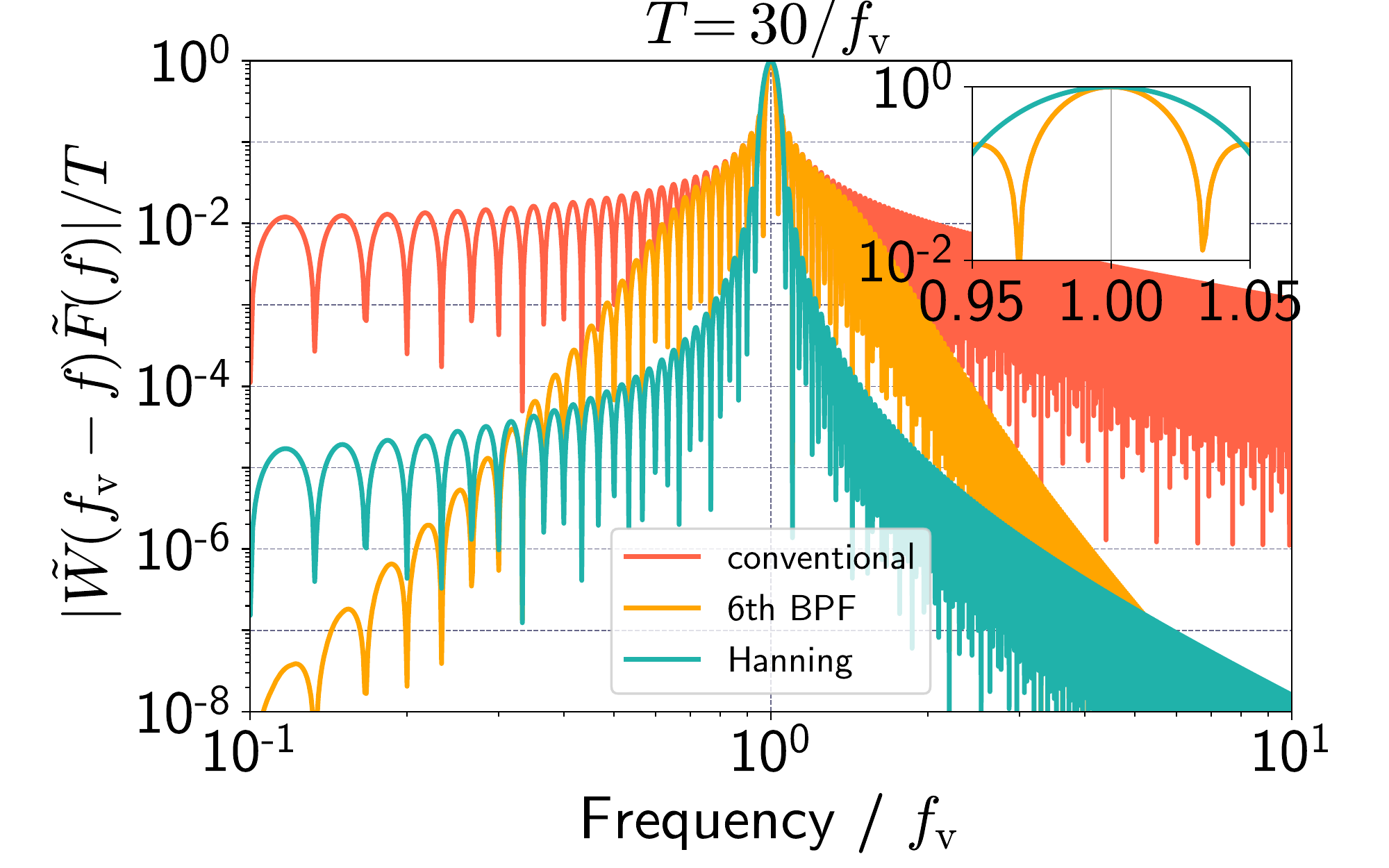}
		\end{center}
	\end{minipage}\\
	\begin{minipage}{1\hsize}
		\begin{center}
		\includegraphics[width=10cm]{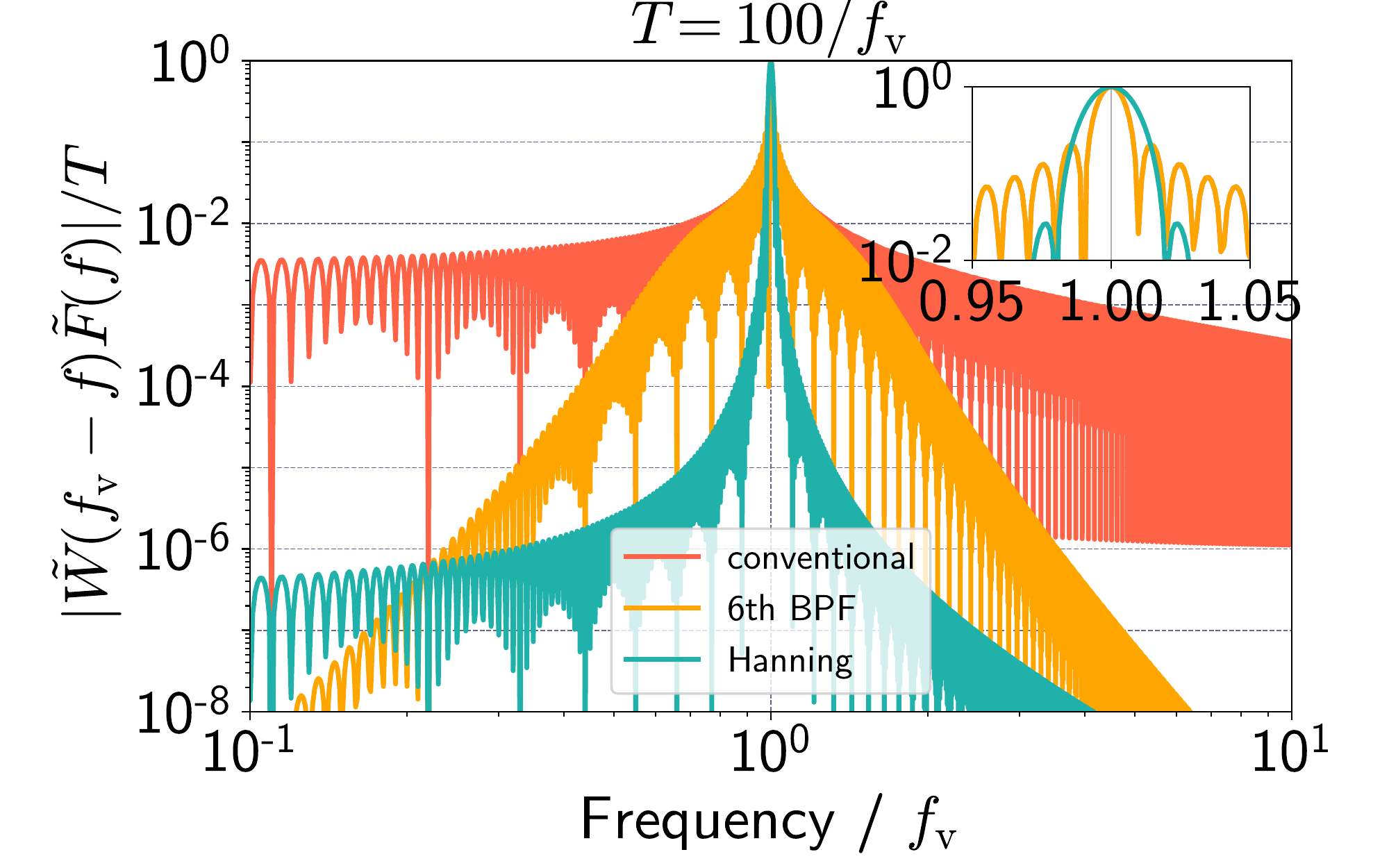}
		\end{center}
	\end{minipage}
	\caption{The amount of the spectral leakage in vibration amplitude estimation from $T=30/f_\mathrm{v}$ (upper) and $100/f_\mathrm{v}$ (lower) data. Three cases using the conventional SAM (red), sixth-order bandpass filtering (orange), and Hanning window (green), are plotted. The enlarged view around $f/f_\mathrm{v}=1$ is shown in the inset.}
	\label{fig:window-filter}
\end{figure}
The effect of the window function is compared with filtering in Fig.~\ref{fig:window-filter}.
The factor $|\tilde{W}(f_\mathrm{v}-f)\tilde{F}(f)|/T$, which determines the amount of the leakage in both the random noise and line noise, is plotted for $T=30/f_\mathrm{v}$ and $100/f_\mathrm{v}$ in three cases: using the rectangular window (conventional SAM), by applying the sixth-order BPF ($Q=1$) with the rectangular window, and using the Hanning window without filtering.
As the figures show, the Hanning window can reduce the leakage more effectively than the BPF around the vibration frequency.
The reduction ratio of the Hanning window in $0.7<f/f_\mathrm{v}<1.5$ is approximately equal to the second-order BPF with $Q\simeq15$ and $50$ for $T=30/f_\mathrm{v}$ and $100/f_\mathrm{v}$, respectively.
In such cases, the long measurement time is wasted due to the settling time, as shown in Fig.~\ref{fig:filter.settling}, while changing the window function does not require discarding the data.
On the other hand, the BPF has a better reduction ratio at frequencies away from $f_\mathrm{v}$, although the details depend on the design of the filter.
The window functions also increase the leakage from a very close frequency to $f_\mathrm{v}$ compared to the rectangular window, as shown in the inset in Fig.~\ref{fig:window-filter}.
In other words, the frequency resolution is compromised by changing the window function.
Therefore, when there is a large low-frequency noise or the calibration frequency is close to a large line noise (e.g., calibration at 49~Hz when there is a large power supply line noise at 50~Hz), bandpass filtering with the rectangular window can be a better choice.
In the scope of vibration frequency determination, the frequency resolution of signal processing is usually not important in vibration calibration because the frequency is accurately controlled based on reference frequency standards.

\subsection{Numerical differentiation}\label{sec:differentiation}
For the common vibration noise $n_x$, the factor $((2\pi f)^2-(2\pi f_\mathrm{v})^2)$ is included as in Eq.~(\ref{eq:uScal.common}).
Similarly, the uncertainty from the common line noise, Eq.~(\ref{eq:uScal.line.common}), contains the factor $|f_\mathrm{v}^2-f_{l,x}^2|$.
These factors originate from the difference of the measured physical quantity between the sensor and the reference interferometer.
The accelerometer measures the acceleration, while the interferometer measures the displacement.
In the conventional SAM for the reference signal, the displacement amplitude is estimated from $x_\mathrm{r}$; then, the acceleration amplitude is calculated by multiplying $(2\pi f_\mathrm{v})^2$.
Although this gives an accurate estimation for a purely sinusoidal wave, the background noise degrades calibration accuracy.

A natural way to avoid such an effect is converting the reference displacement signal into acceleration by numerical differentiation.
Then, the amplitude of the reference signal is estimated in the unit of acceleration.
This process ideally eliminates the factor $((2\pi f)^2-(2\pi f_\mathrm{v})^2)$ in Eq.~(\ref{eq:uScal.common}) and $|f_\mathrm{v}^2-f_{l,x}^2|$ in Eq.~(\ref{eq:uScal.line.common}).
Therefore, the common noise sources $n_x$ and $l_x$ no longer interfere with the calibration.
A similar process is also applicable to the velocity sensor (e.g., seismometer) by changing the number of differentiation.
The important point is to align the units of the signals.

The differentiation process of the displacement signal relatively enlarges the high-frequency noise in the time domain.
Additional processing such as low-pass filtering or changing the window function is required to avoid increasing noise contributions.
On the other hand, the low-frequency noise, which is often included as a drift component in the reference interferometer signal, is reduced because the differentiation process works similarly to a second-order high-pass filter with infinite cutoff frequency.

\subsection{Adjusting data length}\label{sec:adjustT}
The line noise contribution can be eliminated by setting proper data length $T$ so that both $f_\mathrm{v}-f_l$ and $f_\mathrm{v}+f_l$ are at the zero points of $\tilde{W}(f)$.
The Fourier transform of the rectangular window, $\tilde{W}_\mathrm{r}(f)$, is equal to zero at $fT=N$ ($N$: integer). 
Recalling that $f_\mathrm{v}T$ is set to an integer to eliminate the harmonics, as mentioned in Section~\ref{sec:SAM.line}, $f_l T$ also needs to be an integer.
Consequently, if $T$ is an integer multiple of the inverse of the greatest common divisor of $f_\mathrm{v}$ and $f_l$, both the harmonics and line noise can be eliminated.
For example, in the case of $f_l=50$~Hz and $f_\mathrm{v}=49.2$~Hz, the greatest common divisor frequency is 0.4~Hz; hence $T$ should be the multiple of 2.5~s.
The condition is similar for the Hanning window, although it has two fewer zero points around $f_l$ than the rectangular window, as shown in Fig.~\ref{fig:window-filter}.
In usual vibration calibration, $f_\mathrm{v}$ is selected from the one-third octave band specified in ISO 266:1997, and the frequencies are rounded to 0.5~Hz increment around the main line noise frequency (50~Hz).
In such a case, $T$ should be multiple of 2~s.
If there are large line noises at several frequencies, it will be good to set $T$ so that the closest line noise to $f_\mathrm{v}$ is eliminated.

One of the possible drawbacks of adjusting the data length is that the measurement time can be somewhat longer in some cases due to the limited choice of $T$.
If the elimination is impossible within a reasonable measurement time, the line noise contribution needs to be reduced using other methods proposed in Section~\ref{sec:filter} to \ref{sec:differentiation}.
Note that the adjustment is not necessary at every frequency because the line noise contribution becomes a problem only around the line frequency.
Therefore, the total measurement time of calibration does not substantially increase over a wide frequency range.

\subsection{Simulation of proposed methods}
\begin{figure}
	\begin{minipage}{1\hsize}
		\begin{center}
		\includegraphics[width=12cm]{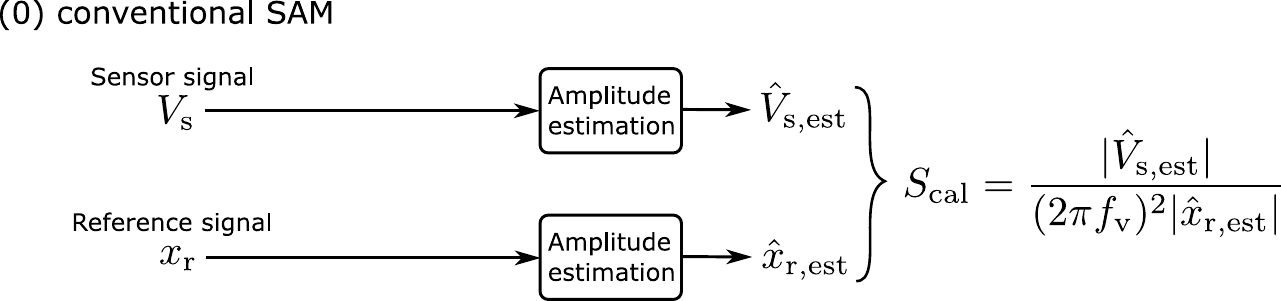}
		\end{center}
		\vspace{5mm}
	\end{minipage}\\
	\begin{minipage}{1\hsize}
		\begin{center}
		\includegraphics[width=12cm]{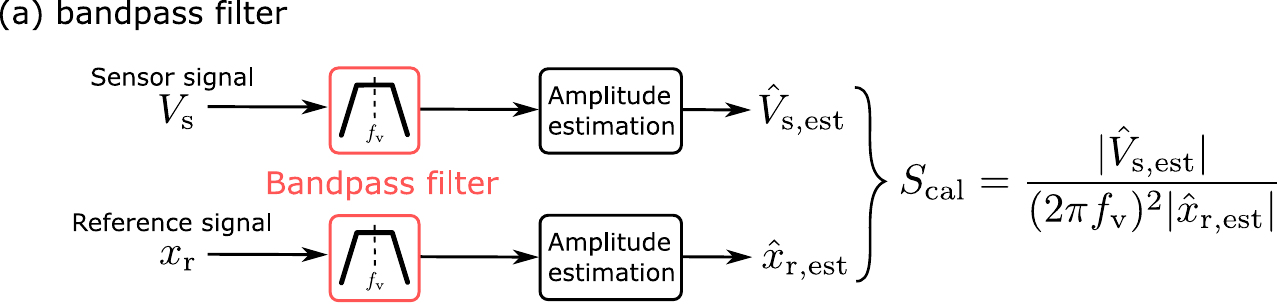}
		\end{center}
		\vspace{5mm}
	\end{minipage}\\
	\begin{minipage}{1\hsize}
		\begin{center}
		\includegraphics[width=12cm]{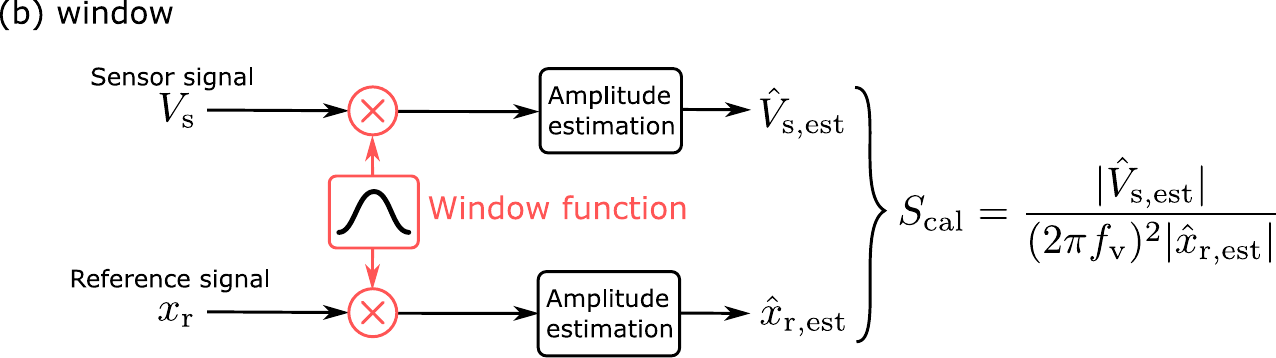}
		\end{center}
		\vspace{5mm}
	\end{minipage}\\ 
	\begin{minipage}{1\hsize}
		\begin{center}
		\includegraphics[width=11cm]{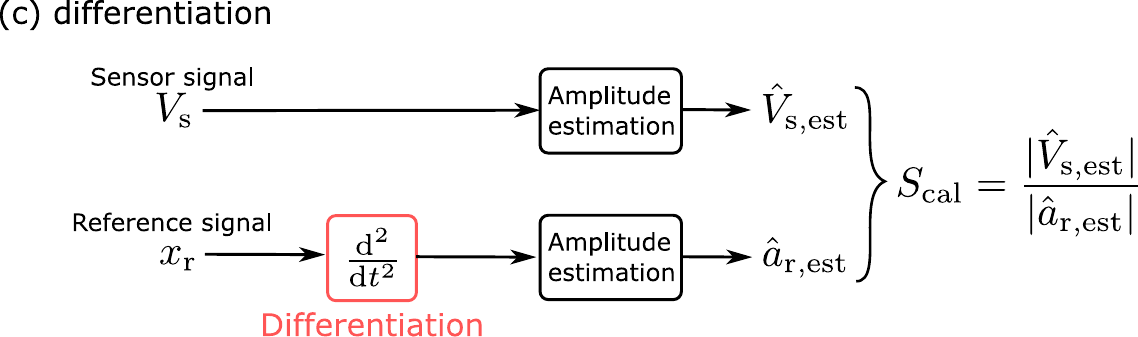}
		\end{center}
	\end{minipage}
\caption{Summary of the signal processing methods for sensitivity calibration including (0) the conventional SAM and the proposed methods; (a) bandpass filtering, (b) multiplying window function, and (c) differentiation of the reference displacement signal.}
\label{fig:signal.processings}
\end{figure}
So far, we have proposed three signal processing methods for sensitivity estimation, which are summarized in Fig~\ref{fig:signal.processings}.
Proper selection of the data length $T$ to remove the line noise contribution was also discussed.
To confirm their effectiveness, a simulation of amplitude estimation was performed for the two types of random noise discussed in Section~\ref{sec:SAM.random} and the line noise mentioned in Section~\ref{sec:SAM.line}.

\subsubsection{Reduction of independent random noise\\}
\begin{figure}
	\begin{minipage}{1\hsize}
		\begin{center}
		\includegraphics[width=10cm]{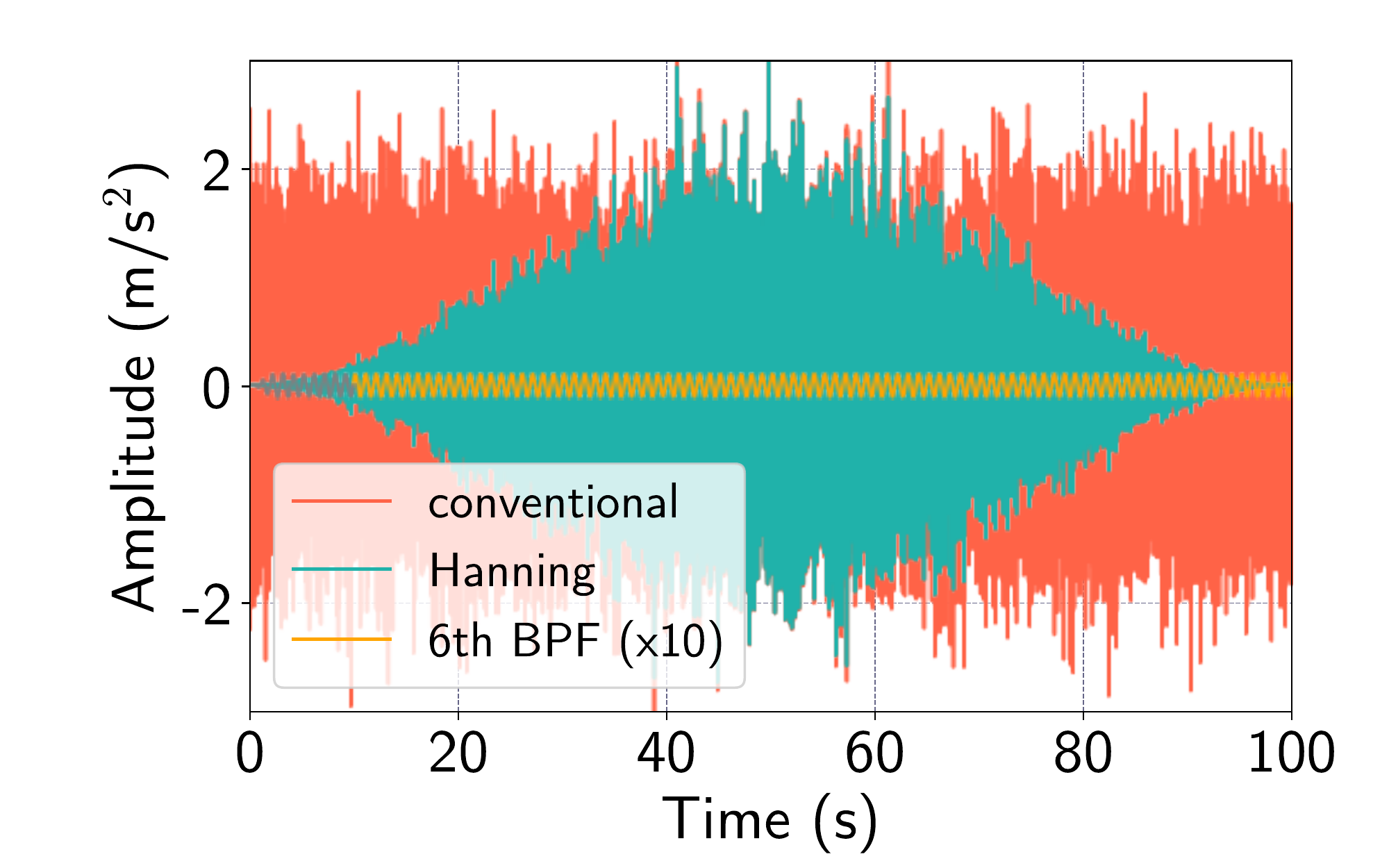}
		\end{center}
	\end{minipage}\\
	\begin{minipage}{1\hsize}
		\begin{center}
		\includegraphics[width=10cm]{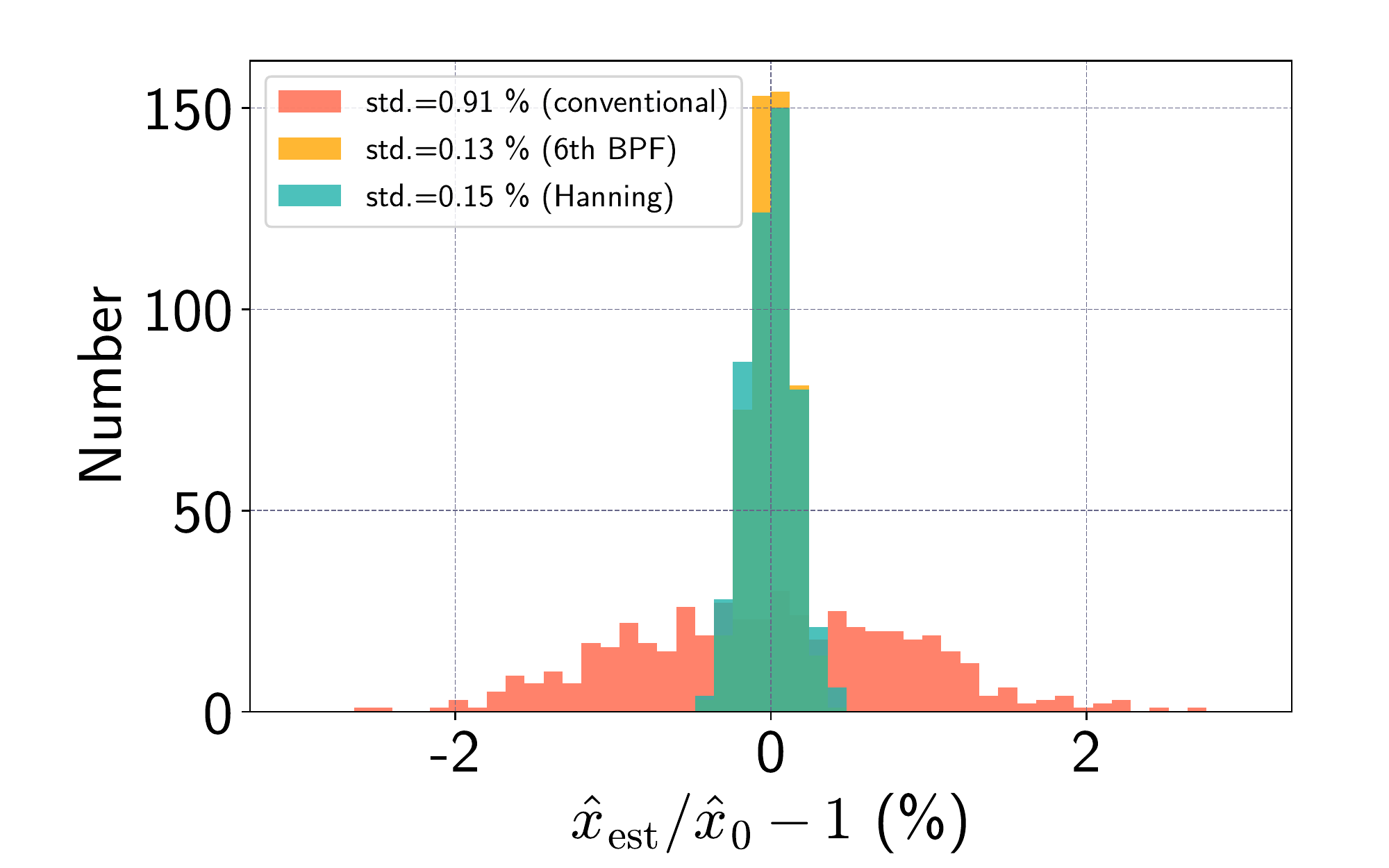}
		\end{center}
	\end{minipage}
\caption{Waveforms used in the simulation ($f_\mathrm{v}=1$~Hz, $(2\pi f_\mathrm{v})^2\hat{x}_0=1$~m/s$^2$) (top). Three cases using the conventional SAM (red), sixth-order bandpass filtering (orange), and Hanning window (green), are plotted. For the second case, the data of the first ten vibration cycles (grey colored) is not used for amplitude estimation. Histograms of the estimated amplitudes for the three cases (bottom).}
\label{fig:nr.sim.reduce}
\end{figure}
First, the independent noise sources, $n_\mathrm{s}$ and $n_\mathrm{r}$, were investigated.
The additional signal processing methods proposed in Section~\ref{sec:filter} and \ref{sec:window} were effective in this case and corresponded to (a) and (b) in Fig.~\ref{fig:signal.processings}, respectively.
Here, the amplitude estimation uncertainty of the single signal ($V_\mathrm{s}$ or $x_\mathrm{r}$) was simulated for simplicity.
Using the results, the sensitivity calibration uncertainty can be calculated based on Eq.~(\ref{eq:uScal.random}).
A simulation similar to that in Section~\ref{sec:indep.random} was performed; the amplitude estimation of sinusoidal waveform with $(2\pi f_\mathrm{v})^2\hat{x}_0=1$~m/s$^2$ under the background noise as the right column case of Fig.~\ref{fig:random.simulation}.
In this case, the standard deviation of the estimated amplitude increased to 0.83~\% based on the conventional SAM due to the noise at $f\neq f_\mathrm{v}$, while the S/N at the vibration frequency was 0.1~\%.
Here, the sixth-order Butterworth BPF with $Q=1$ and the Hanning window were applied to the noisy waveform, and the amplitude was estimated from the processed waveforms.
The examples of the processed waveforms are shown in Fig.~\ref{fig:nr.sim.reduce} (upper figure).
The data of the first ten vibration cycles were discarded for the filtered waveform, as explained in Section~\ref{sec:filter}.
The simulation was repeated 300 times for the randomly generated noise, and the histogram of the amplitude estimation error is shown in Fig.~\ref{fig:nr.sim.reduce} (lower figure).
For the conventional SAM, BPF, and Hanning window, the simulated standard deviations were 0.91~\%, 0.13~\%, and 0.15~\%, and the expected deviations from Eq.~(\ref{eq:std.n}) were 0.83~\%, 0.13~\%, and 0.15~\%, respectively.
As expected, the BPF and Hanning window reduce the background noise contributions at $f\neq f_\mathrm{v}$.
Although they do not achieve an S/N of 0.1~\% at $f_\mathrm{v}$ due to the leakage around $f_\mathrm{v}$, the difference is sufficiently small.
Consequently, the calibration uncertainty is also reduced because Eq.~(\ref{eq:uScal.random}) is the sum of the amplitude estimation uncertainties of $V_\mathrm{s}$ and $x_\mathrm{r}$. 

In any cases shown in Fig.~\ref{fig:nr.sim.reduce}, the amplitude estimation errors are distributed around zero, which indicates that there is no bias due to the filter or window.
The correction of the filter gain is necessary, depending on the choice of the filter.
To avoid the systematic effects on sensitivity calibration, applying the same process on both the sensor and reference signals is recommended to cancel out the effect.

\subsubsection{Reduction of common random noise\\} \label{sec:recuce.common.random}
\begin{figure}
	\begin{center}
	\includegraphics[width=10cm]{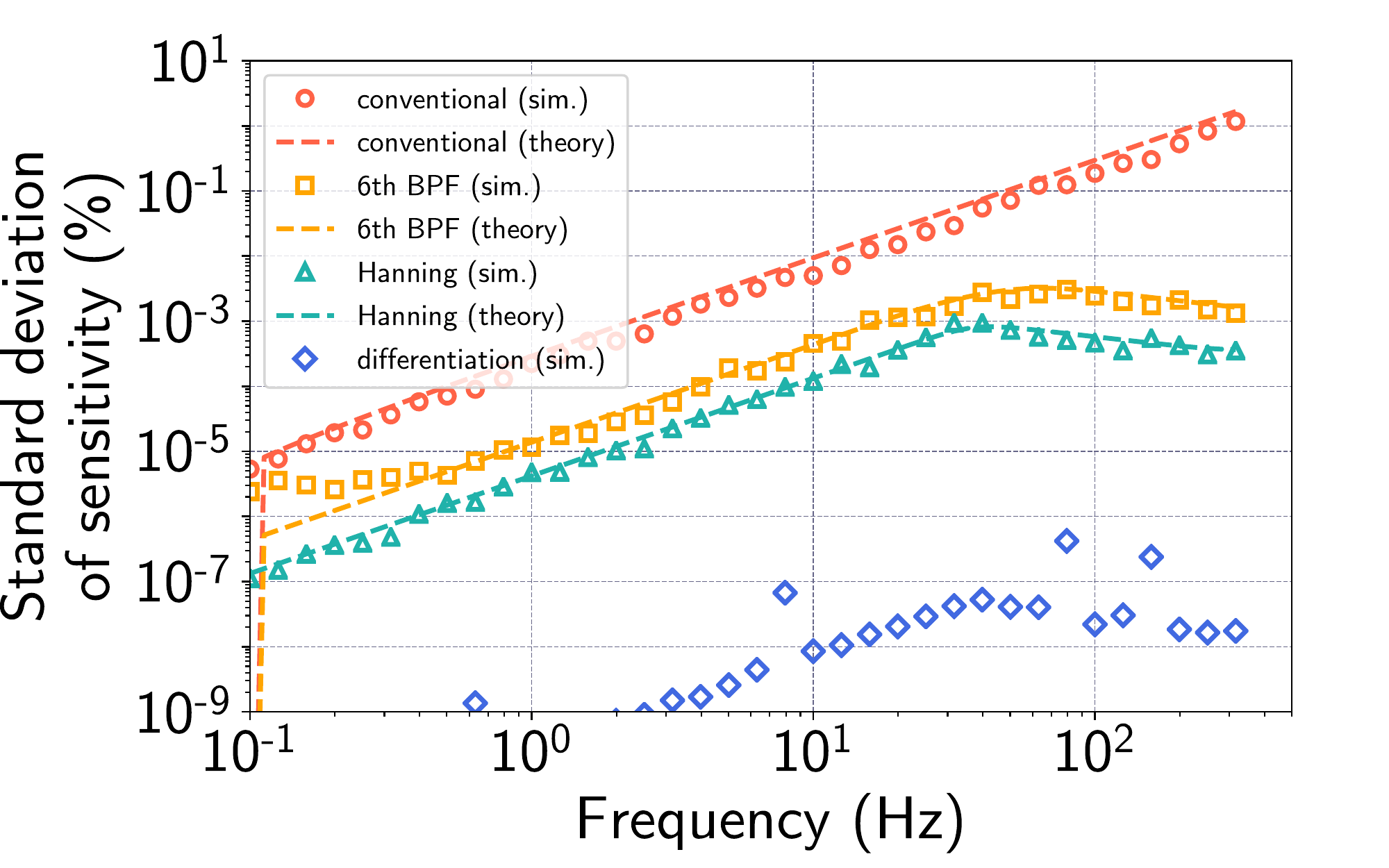}
	\caption{Standard deviation of the estimated sensitivity under the random vibration noise $n_x$ (Fig.~\ref{fig:common.simulation}) for four cases; using the conventional SAM (red), sixth-order bandpass filter (orange), Hanning window (green), and numerical differentiation (blue). The simulation results are shown with the circles, and the theoretical expectations from Eq.~(\ref{eq:uScal.common}) are plotted with the dashed lines. The theoretical expectation for the numerical differentiation is not shown because it is zero.}
	\label{fig:nx.sim.reduce}
	\end{center}
\end{figure}
Second, the common vibration noise, $n_x$, was investigated.
The signal processing methods proposed in Section~\ref{sec:filter}, \ref{sec:window}, and \ref{sec:differentiation} were effective in this case.
They correspond to (a), (b), and (c) in Fig.~\ref{fig:signal.processings}, respectively.
The noise spectrum in Fig.~\ref{fig:common.simulation} and the excitation amplitude of $(2\pi f_\mathrm{v})^2\hat{x}_0=1$~m/s$^2$ were used for the simulation.
In (a) and (b), the same BPF and Hanning window as the previous simulation (Fig.~\ref{fig:nr.sim.reduce}) were used.
The amplitudes were estimated using the processed data, and the sensitivity was calculated from Eq.~(\ref{eq:Scal}).
In (c), the reference displacement data $\{x_{\mathrm{r}, n}\}$ ($n=0, 1, ..., N-1$) was converted to the reference acceleration $a_{\mathrm{r}, n}$:
\begin{equation}
	a_{\mathrm{r}, n} = ( x_{\mathrm{r}, n+1} - 2 x_{\mathrm{r}, n} + x_{\mathrm{r}, n-1} ) f_\mathrm{s}^2,
\label{eq:differentiation}
\end{equation}
where $f_\mathrm{s}$ is the sampling frequency.
Then, the amplitudes were estimated using $\{v_{\mathrm{s}, n}\}$ and $\{a_{\mathrm{r}, n}\}$, and the sensitivity was calculated as $S_\mathrm{cal}=|\hat{V}_\mathrm{s, est}/\hat{a}_\mathrm{r, est}|$ instead of Eq.~(\ref{eq:Scal}).
These simulations are repeated 20 times for the randomly generated noise at vibration frequencies $0.1~\mathrm{Hz} < f_\mathrm{v}<300~\mathrm{Hz}$.
The simulation results and theoretical expectations from Eq.~(\ref{eq:uScal.common}) are shown in Fig.~\ref{fig:nx.sim.reduce}.
Both the BPF and Hanning window reduced the relative standard uncertainty of sensitivity from 0.3~\% to 0.01~\% and 0.001~\%, respectively, around 100~Hz.
The reduction effect agreed well with the theory.
Numerical differentiation is much more effective, as shown in the figure.
Therefore, it is recommended to adopt the differentiation process in accelerometer calibration when the background vibration noise is large.

Nevertheless, the systematic error of numerical differentiation requires attention.
The transfer function of Eq.~(\ref{eq:differentiation}) from the displacement to acceleration is given by
\begin{equation}
	( e^{2\pi i f /f_\mathrm{s}} - 2 + e^{-2\pi i f /f_\mathrm{s}} ) f_\mathrm{s}^2 = -(2\pi f)^2 \left( \mathrm{sinc}(f/f_\mathrm{s}) \right)^2,
\end{equation}
while the second-order derivative of continuous signal is given by $-(2\pi f)^2$.
The factor $\left( \mathrm{sinc}(f_\mathrm{v}/f_\mathrm{s}) \right)^2$ needs to be corrected when numerical differentiation is applied.
Note that the factor differs from 1 by less than 0.02~\% for $f_\mathrm{v}<0.01f_\mathrm{s}$.
If the sampling frequency is sufficiently faster than the vibration frequency, typically by 100 times, the error is ignorable.

\subsubsection{Reduction of independent line noise\\}
\begin{figure}
	\begin{center}
	\includegraphics[width=10cm]{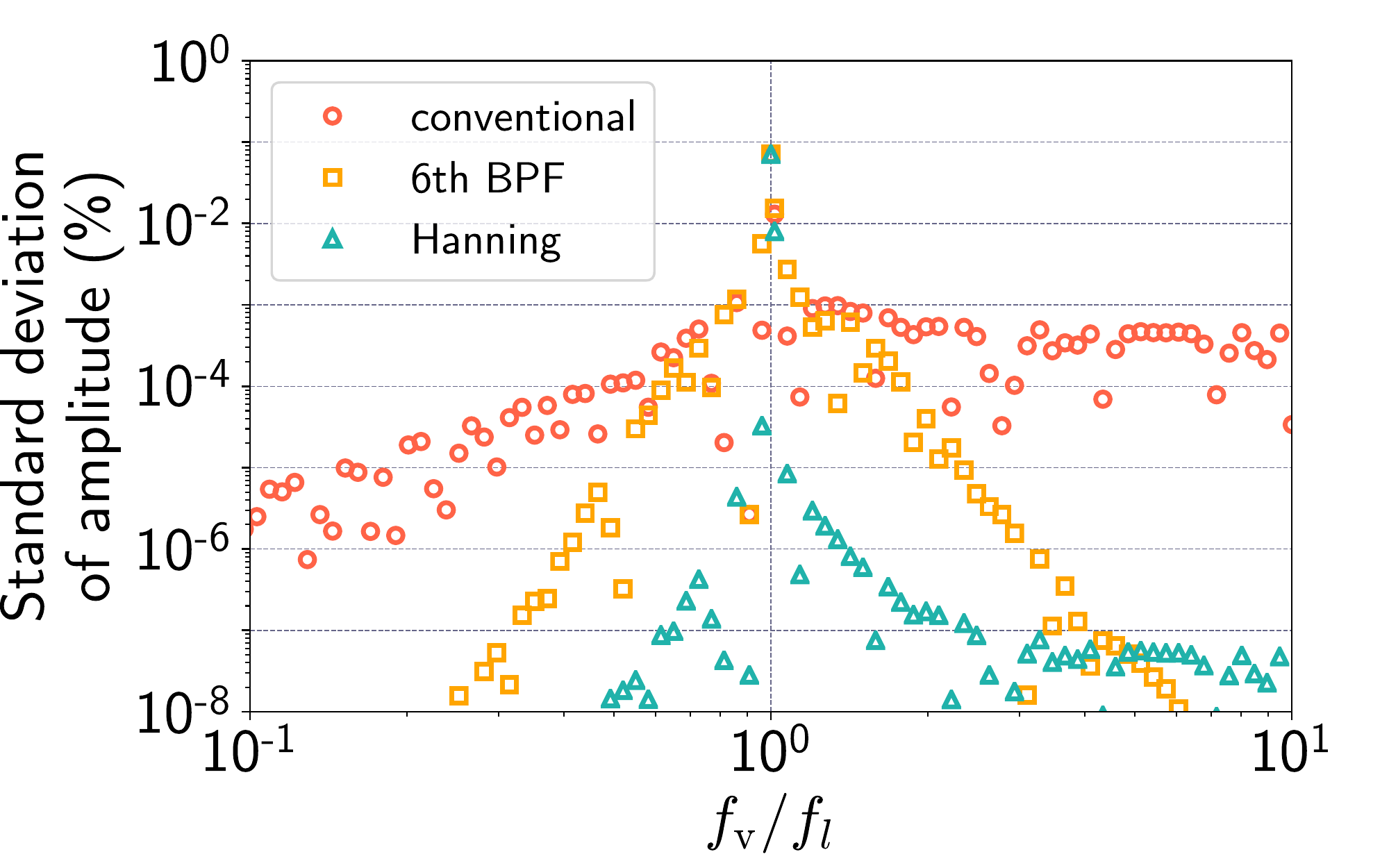}
	\end{center}
\caption{Uncertainty simulation for the line noise with $\hat{l}/\hat{x}_0=10^{-3}$. Three cases using the conventional SAM (red circle), sixth-order bandpass filtering (orange square), and Hanning window (green triangle), are plotted.}
\label{fig:linenoise.50Hz.reduce}
\end{figure}
The same methods applied to the random noise are also effective for the line noise.
For the independent line noise, $l_\mathrm{s}$ and $l_\mathrm{r}$, the filter and window function are useful.
They correspond to (a) and (b) in Fig.~\ref{fig:signal.processings}, respectively.
The effects of the BPF and Hanning window on the independent line noise were simulated here.
The signal $x(t)=x_0(t)+l(t)$ was filtered by the sixth-order BPF with $Q=1$ or windowed using the hannin window.
The amplitude estimation uncertainties were simulated using the same process described in Section~\ref{sec:SAM.line}.
The result is shown in Fig.~\ref{fig:linenoise.50Hz.reduce}.
As Eq.~(\ref{eq:ux.line.approx}) indicates, the reduction effect is similar to the frequency dependence of $|\tilde{W}(f_\mathrm{v}-f)\tilde{F}(f)|/T$, shown in Fig.~\ref{fig:window-filter}.
The Hanning window is more effective than the BPF over the simulated frequency range, where the line noise has a large uncertainty contribution.

\begin{figure}
	\begin{center}
	\includegraphics[width=10cm]{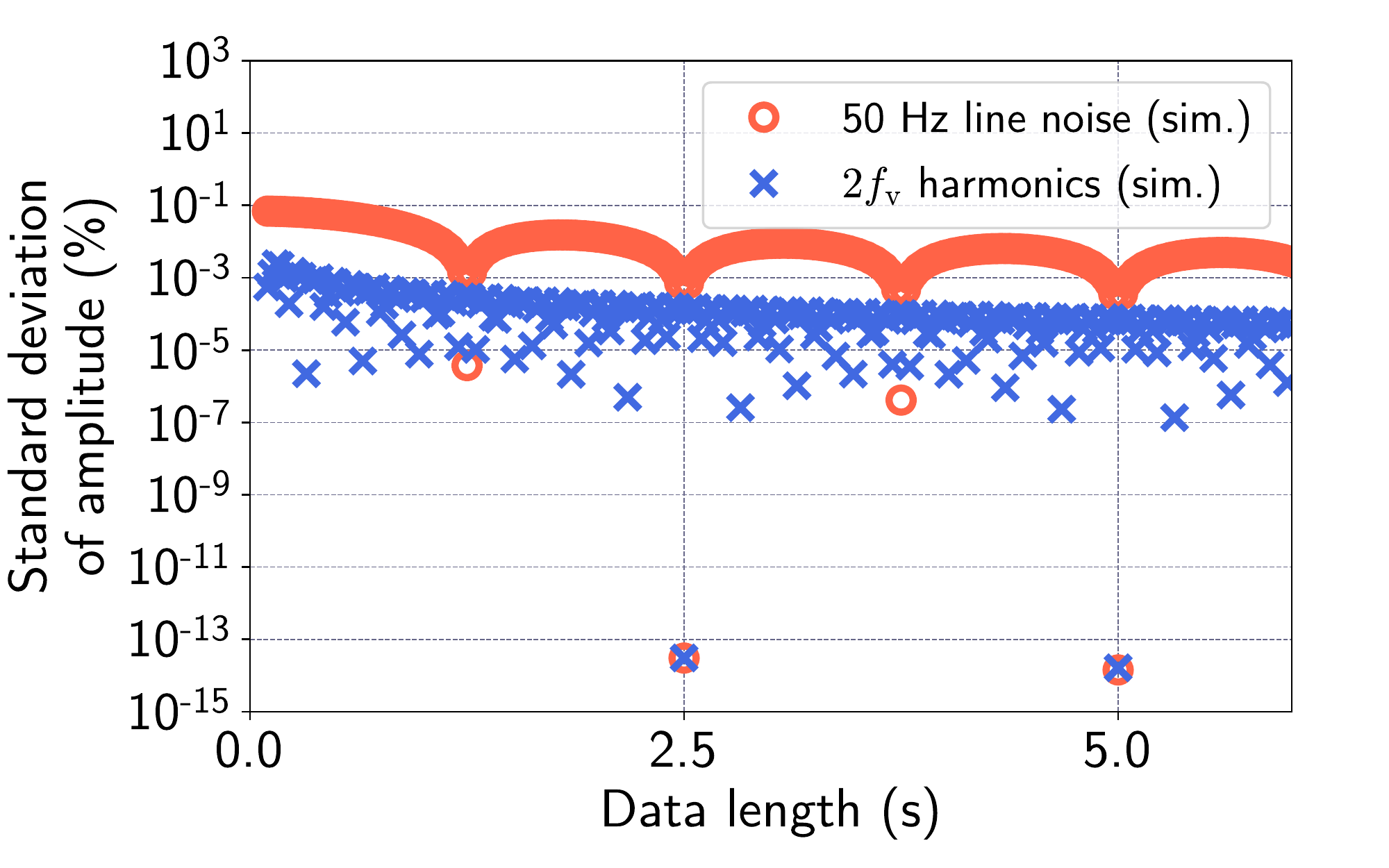}
	\caption{Simulated amplitude estimation uncertainty at $f_\mathrm{v}=49.2$~Hz under the line noise at $f_l=50$~Hz (red) and harmonics at $f_l=2f_\mathrm{v}$ for different data length $T$. The line amplitude is $\hat{l}/\hat{x}_0=10^{-3}$ at each frequency.}
	\label{fig:line.adjustT}
	\end{center}
\end{figure}
To confirm that the proper choice of $T$ can eliminate the line noise contribution, as discussed in \ref{sec:adjustT}, the dependence on the data length was also simulated.
As an example, the calibration frequency was fixed at $f_\mathrm{v}=49.2$~Hz, and the data length was varied changed from 0.1~s to 6~s.
The amplitude estimation uncertainty for $x(t)=x_0(t)+l(t)$ was simulated using the same method described in Section~\ref{sec:indep.line} (using the rectangular window withoug a filter).
For $l(t)$, both the line noise at fixed frequency $f_l=50$~Hz and harmonics at $f_l=2f_\mathrm{v}$ were used for the simulation.
The results are shown in Fig.~\ref{fig:line.adjustT}.
As discussed in Section~\ref{sec:adjustT}, both uncertainties from the line noise and harmonics reach almost zero (limited by numerical computation error) when $T$ is multiple of 2.5~s, which is the inverse of the greatest common divisor of $f_\mathrm{v}$ and $f_l$.

\subsubsection{Reduction of common line noise\\}
\begin{figure}
	\begin{center}
	\includegraphics[width=10cm]{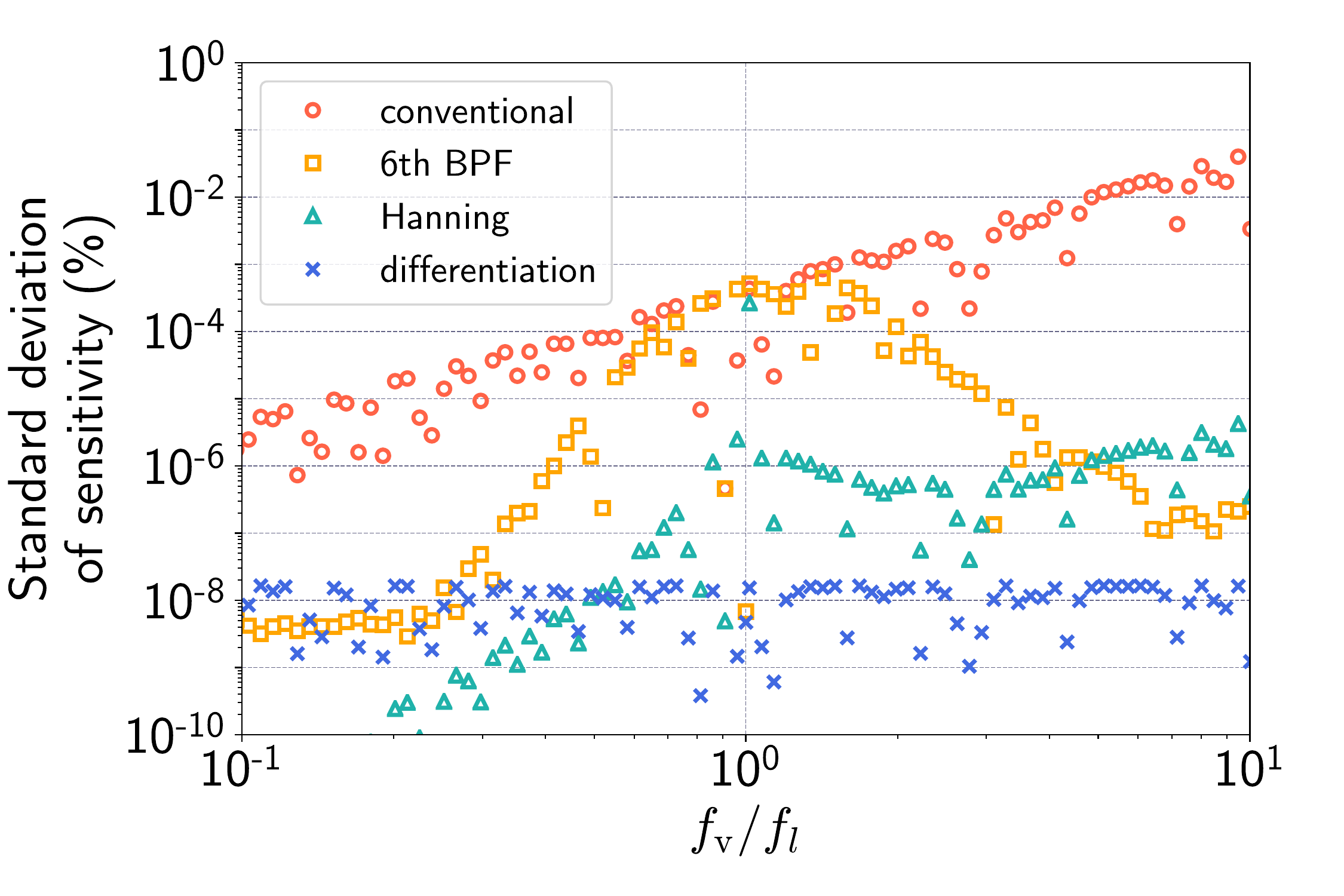}
	\end{center}
\caption{Sensitivity estimation uncertainty due to the common line noise $l_x$ with $\hat{\ddot{l}}_x/\hat{\ddot{x}}_0=10^{-3}$. The simulated uncertainties for four cases using the conventional SAM (red circle), sixth-order bandpass filter (orange square), Hanning window (green triangle), and numerical differentiation (blue cross), are plotted.}
\label{fig:linenoise.50Hz.common.reduce}
\end{figure}
For the common line noise, $l_x$, numerical differentiation is effective.
The uncertainties with the conventional SAM, sixth-order BPF, Hanning window, and numerical differentiation were simulated and plotted in Fig.~\ref{fig:linenoise.50Hz.common.reduce}.The simulation was performed under the same conditions, as described in Section~\ref{sec:common.line}.
The same filter, window, and differentiation process described in Section~\ref{sec:recuce.common.random} were applied.
As expected, those signal process modifications reduce the common line noise contributions.
Especially, the numerical differentiation eliminates the common line noise contribution over a wide frequency range.

Note that the proper choice of $T$ is also effective for reducing the common line noise.
Therefore, the modification of signal processing is not necessary for line noise reduction, if it is already eliminated by the choice of $T$.

\section{Application to the accelerometer calibration in NMIJ}	\label{sec:experiment}
In NMIJ, the calibration system with a small excitation amplitude ($(2\pi f_\mathrm{v})^2\hat{x}_0\simeq 10^{-2}$~m/s$^2$) is under development \cite{calibration.VIS}.
The background noise of the calibration system becomes a significant uncertainty source in micro vibration calibration.
Here, the proposed processing methods are applied to the actual calibration data to reduce uncertainty.
The calibration system is shown in Fig.~\ref{fig:system}.
A servo accelerometer JA-5V (Japan Aviation Electronics Industry, Ltd., $S\simeq0.1$~V/(m/s$^2$) (nominal)) was used as the calibration target.
Since the calibration uncertainty discussed in this paper mainly affects repeatability, the calibration was repeated 5 times at frequencies from 0.4~Hz to 500~Hz, and their standard deviations were measured at each frequency.
The results were compared with the theoretical limits due to the background random noise of $V_\mathrm{s}$ and $x_\mathrm{r}$ (as Eq.~(\ref{eq:std.cns.approx})) in our system.
The effect of the line noise was limited and not large compared to the random noise; hence, we mainly discuss the random noise in this section.
About the independent background noise, $n_\mathrm{s}$ includes the self-noise of the accelerometer and the noise of the signal acquisition system, and $n_\mathrm{r}$ includes the self-noise of the laser interferometer, seismic vibration noise, and signal acquisition system noise.
The common background noise $n_x$ is shown in Fig.~\ref{fig:common.simulation}.
Relatively large low-frequency drift, which also behaves as the common background noise, was applied above 30~Hz for averaging the cyclic error of the interferometer.

\begin{figure}
	\begin{minipage}{1\hsize}
		\begin{center}
		\includegraphics[width=12cm]{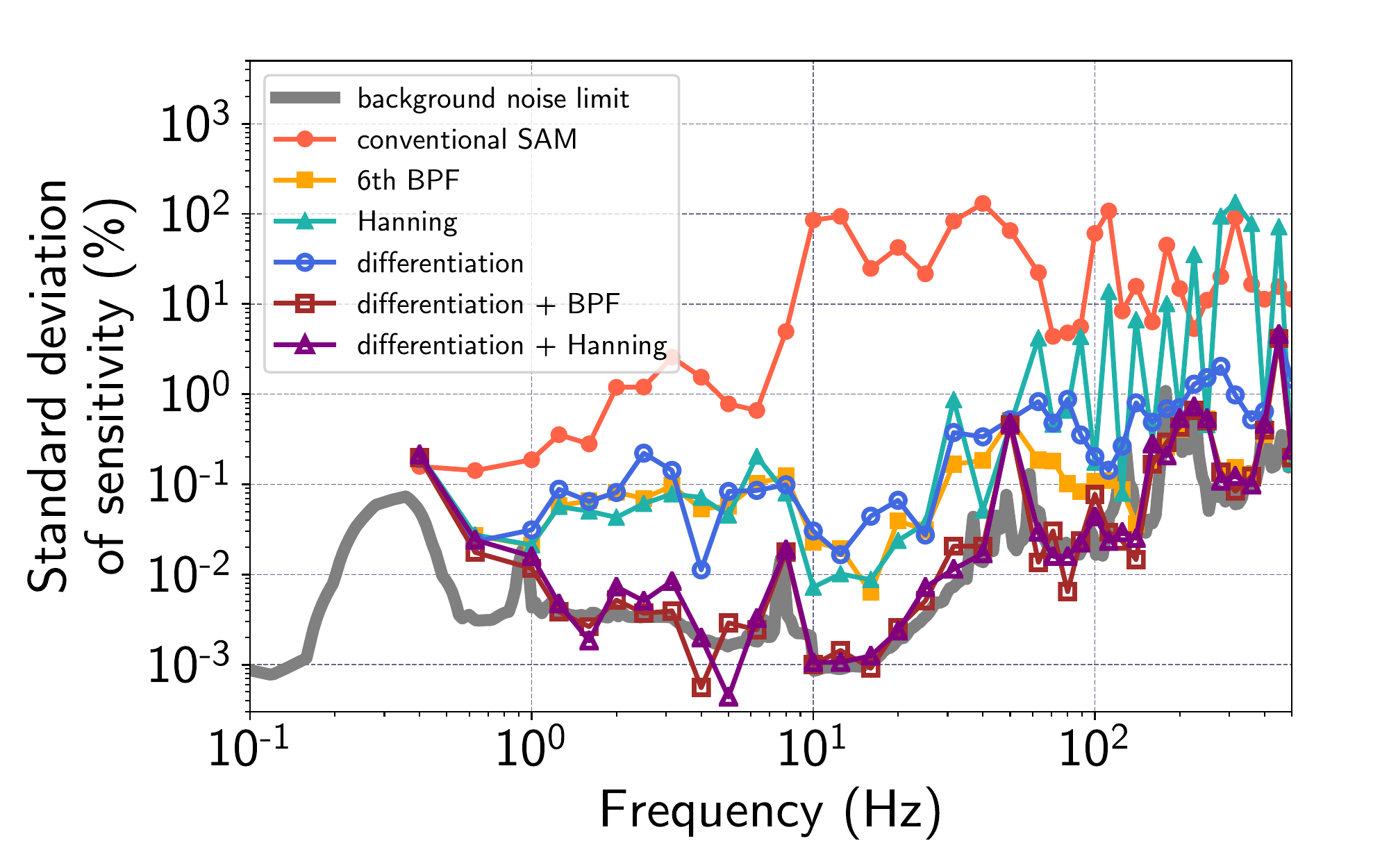}
		\end{center}
	\end{minipage}\\
	\begin{minipage}{1\hsize}
		\begin{center}
		\includegraphics[width=12cm]{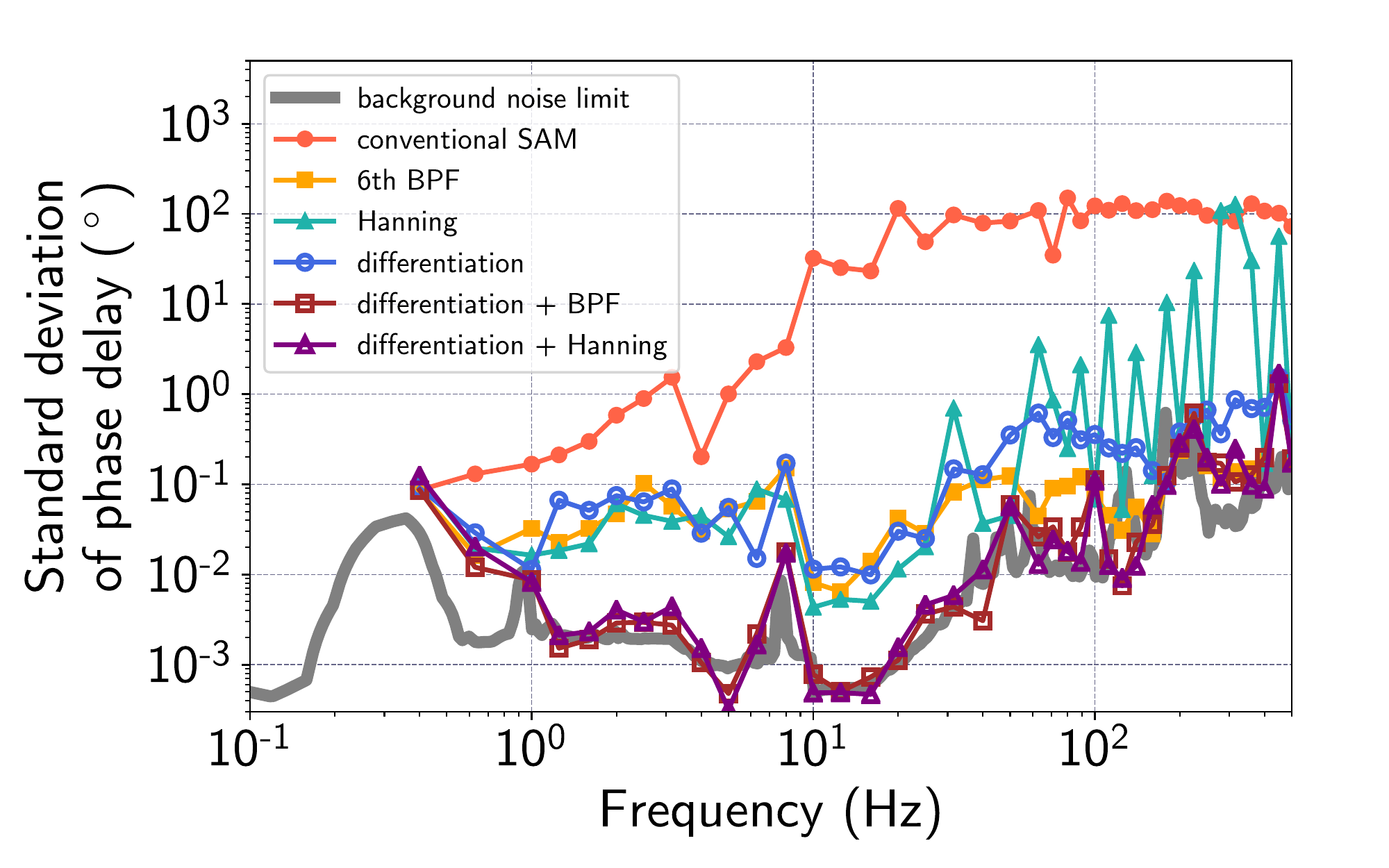}
		\end{center}
	\end{minipage}
\caption{Standard deviations of the calibrated sensitivities (upper) and phase delays (lower) repeated 5 times with $(2\pi f_\mathrm{v})^2\hat{x}_0\simeq 10^{-2}$~m/s$^2$ using the calibration system in NMIJ. The results using different signal processing methods are plotted: the conventional SAM (red solid circle), sixth-order BPF (orange solid square), Hanning window (green solid triangle), numerical differentiation for $x_\mathrm{r}$ (blue open circle), differentiation and BPF (brown open square), and differentiation and Hanning window (purple open triangle). The theoretical limit due to the background noise of the system is shown by the solid grey line.}
\label{fig:actual}
\end{figure}
The experimentally obtained repeatabilities with different signal processing methods are shown in Fig.~\ref{fig:actual}.
The sixth-order BPF, Hanning window, numerical differentiation (Eq.~(\ref{eq:differentiation})), and their combinations were adopted.
With the conventional SAM without any additional processing methods, the repeatability uncertainty was about 100~\% over a broad frequency range. 
The calibration results were almost meaningless in such a large uncertainty.
By using the BPF, Hanning window, or numerical differentiation, the repeatability was improved by up to four orders of magnitude.
Since numerical differentiation could reduce uncertainty, the overall repeatability was thought to be limited by the common vibration noise $n_x$ or low-frequency noise in the reference displacement signal $n_\mathrm{r}$.
At high frequencies above 50~Hz, the noise reduction is insufficient only in the case of the Hanning window because the low-frequency drift is not suppressed enough.
The BPF is a better choice in such a case.
The combinations of numerical differentiation with BPF or Hanning window were more effective than the single processing.
The results with the combined processing methods are also shown in Fig.~\ref{fig:actual}, which achieved the theoretical limit by the calibration system noise.
As expected from Eq.~(\ref{eq:std.arg}), the reduction effects are almost the same for the phase delay.

Our experiment proved that the signal processing methods proposed in this paper are useful in actual calibration systems.
The results agreed with the theoretically expected reduction ability; therefore, the calculations shown in Section~\ref{sec:calculation} and \ref{sec:reduce} can be used for optimizing signal processing.
For example, in the case of the accelerometer calibration in NMIJ, the common vibration noise had a dominant uncertainty contribution.
The independent background noise bottomed out near 10~Hz and increased on the low- and high-frequency side; hence, the calibration around 10~Hz was affected by the leakage from the low- and high-frequency ranges.
Since both the common and independent noises needed to be reduced, the combination of numerical differentiation and the BPF or Hanning window was required.
If the common vibration noise is small, numerical differentiation is not necessary.
The spectral leakage may be insignificant if the independent background noise has a flat spectrum; then, the conventional SAM without the BPF or Hanning window may be sufficient. 
Thus, the processing methods need to be combined based on the calibration system noise.

\section{Conclusions}	\label{sec:conclusion}
\begin{figure}
	\begin{minipage}{1\hsize}
		\begin{center}
		\includegraphics[width=13cm]{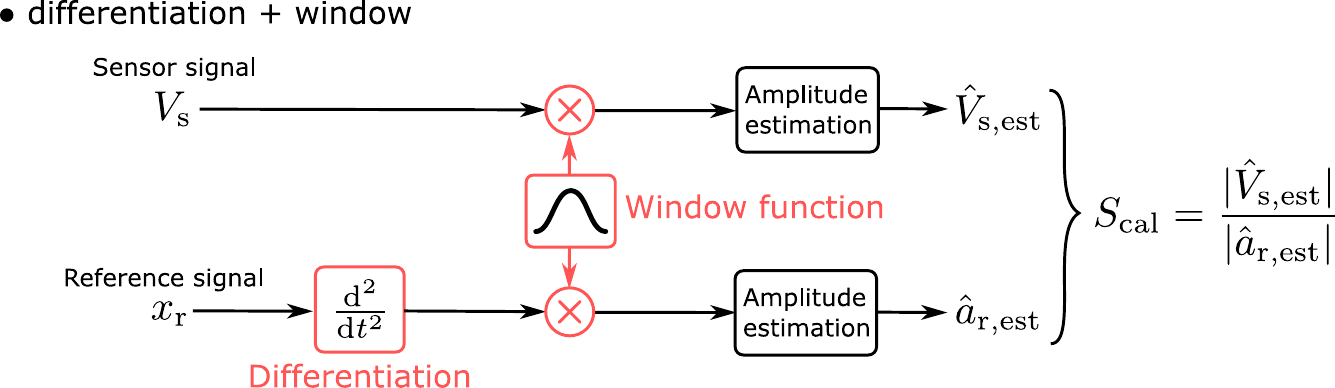}
		\end{center}
		\vspace{5mm}
	\end{minipage}\\
	\begin{minipage}{1\hsize}
		\begin{center}
		\includegraphics[width=13cm]{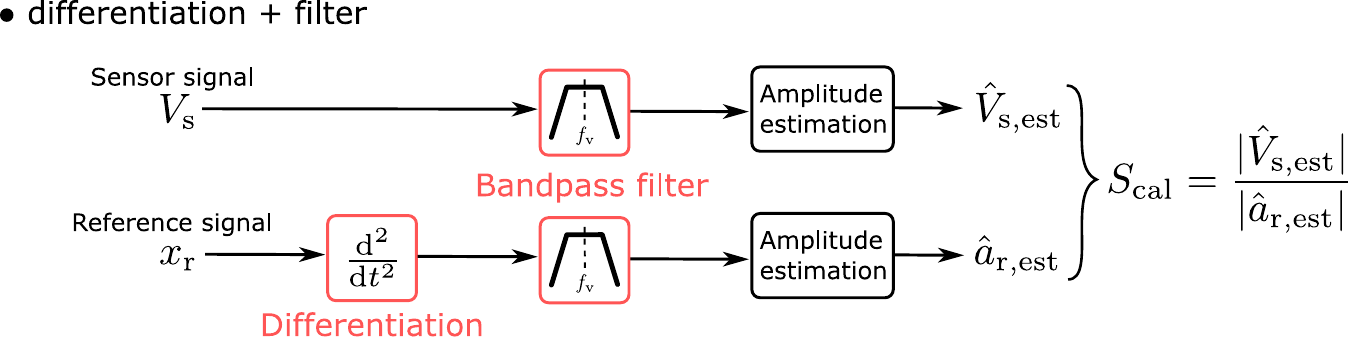}
		\end{center}
		\vspace{5mm}
	\end{minipage}\\
\caption{Combination of the numerical differentiation and the filter or window.}
\label{fig:process.combine}
\end{figure}

We proposed three signal processing methods to modify the conventional SAM specified in ISO16063-11.
The proposed methods include
\begin{itemize}
\item{filtering the signal,}
\item{changing the window function,}
\item{differentiation of the reference displacement signal.}
\end{itemize}
The proper choice of the data length $T$ was also discussed to eliminate the line noise contribution.
Their effect on the reduction of calibration uncertainty was confirmed by both the simulation and experiment.
These results showed that, in accelerometer calibration, using the combination of numerical differentiation and the filter or window before amplitude estimation results in a more robust calibration against the background noise compared to using the conventional SAM.

\begin{table}
\begin{center}
\caption{Summary of the effectiveness ($++$:contribution is eliminated, $+$:contribution is reduced, $-$:ineffective) and advantage/disadvantage of the proposed signal processing methods.}\label{table:summary}
	\begin{tabular}{c|c c c c|p{4cm}|p{4cm}} \hline
	\multirow{2}{*}{method} & \multicolumn{4}{c|}{effect on} & \multirow{2}{*}{advantage} & \multirow{2}{*}{disadvantage} \\	
	 & $n_x$	& $l_x$	& $n_\mathrm{r}$, $n_\mathrm{s}$	& $l_\mathrm{r}$, $l_\mathrm{s}$	&  & 	\\ \hline\hline	
	filter	& $+$	& $+$	& $+$	& $+$	& large reduction ratio at $f$ away from $f_\mathrm{v}$, design is flexible	& data is wasted during the settling time \\ \hline
	window	& $+$	& $+$	& $+$ & $+$	& large reduction ratio at $f$ close to $f_\mathrm{v}$	& low-frequency noise remains \\ \hline
	differentiation	& $++$	& $++$	& $-$	& $-$	& common noise is eliminated, low-frequency reference noise is reduced & high-frequency reference noise is amplified \\ \hline
	adjusting $T$	& $-$	& $++$	& $-$	& $++$	& line noise is eliminated & not always possible within reasonable $T$\\ \hline
	\end{tabular}
\end{center}
\end{table}
Table~\ref{table:summary} summarizes the effectiveness, advantages, and disadvantages of the proposed methods.
The line noise and common noise contributions can be eliminated by adjusting the data length and using numerical differentiation for the reference displacement signal.
If the independent random noise contribution is significant, additional filtering or window function needs to be combined with numerical differentiation.
Fig.~\ref{fig:process.combine} shows the recommended combinations of sensitivity calibration.
The combination of numerical differentiation and Hanning window is sufficient in standard cases.
Although filtering offers flexibility, the data of the settling time needs to be discarded, and instability is a concern for the infinite impulse response filter.
Eqs.~(\ref{eq:uScal.random}), (\ref{eq:uScal.common}), (\ref{eq:uScal.line.indep}), and (\ref{eq:uScal.line.common}) are useful for the optimization based on the system noise characterization.

This work enables accelerometer calibration with a small excitation amplitude relative to the system background noise.
Such calibration is required to confirm the sensitivity linearity of accelerometers used for micro vibration measurements.
Additionally, for accelerometers with high sensitivity, large vibration may not be applied for calibration because they can get saturated with small input vibration. 
In the calibration at a low frequency ($<1$~Hz), which is required for broadband seismometers, the excited acceleration amplitude is limited due to the stroke limit of the exciter.
The reduction of the background noise is necessary in these cases, and proper signal processing is essential to take full advantage of noise reduction.

Although we mainly discussed accelerometer calibration in this work, some of the knowledge obtained  also applies to other fields of dynamic sensor calibration, where sinusoidal signal extraction is required.
When the sensor output signal is compared with a reference signal, their signals should be in unit of the same physical quantity; otherwise, the common background noise affects the estimation of the amplitude ratio.
The demand for measuring small fluctuations is increasing in not only mechanical vibration measurements but also various fields along with the progress of industry.
The proposed signal processing methods for accelerometer calibration can contribute to improving the reliability of those measurements.


\section*{Acknowledgment}
This work was partially based on the results obtained from a project commissioned by the New Energy and Industrial Technology Development Organization (NEDO), Japan.


\section*{References}
\begin{thebibliography}{}
\bibitem{ISHM} Brownjohn J M W 2007
    \newblock{Structural health monitoring of civil infrastructure}, {\it Philosophical Transactions A: Mathematical, Physical, and Engineering Sciences}, {\bf 365}, 589
\bibitem{ESA.handbook} Calvi A and Roy N 2013
	\newblock{Spacecraft mechanical loads analysis handbook}, {\it ESA Requirements and Standards Division, Noordwijk, The Netherlands}
\bibitem{MEMS.Deng} Deng T, Chen D, Wang J, Chen J, Sun Z and Li G 2015
	\newblock{Microelectromechanical systems-based electrochemical seismic sensors with insulating spacers integrated electrodes for planetary exploration}, {\it IEEE Sensors Journal}, {\bf 16}, 3, 650
\bibitem{MEMS.Isobe} Isobe A, Kamada Y, Takubo C, Furubayashi Y, Oshima T, Sakuma N and Sekiguchi T 2020
	\newblock{Design of perforated membrane for low-noise capacitive MEMS accelerometers}, {\it IEEE Sensors Journal}, {\bf 20}, 1184
\bibitem{MEMS.review} Wang C {\it et al.} 2020
	\newblock{Micromachined accelerometers with sub-$\mu$g/$\sqrt{\rm Hz}$ noise floor: A review.}, {\it Sensors} , {\bf 20}, 14, 4054
\bibitem{CCAUV.V-K2} Bruns T, Ripper G P and T\"{a}ubner A 2014
    \newblock{Final report on CIPM key comparison CCAUV.V-K2}, {\it Metrologia}, {\bf 51}, 1A, 09002
\bibitem{CCAUV.V-K3} Qiao S {\it et al.} 2017
    \newblock{Final report of CCAUV.V-K3: Key comparison in the field of acceleration on the complex charge sensitivity}, {\it Metrologia}, {\bf 54}, 1A, 09001
\bibitem{ISO16063-11} International Organization for Standardization
    \newblock{ISO 16063-11:1999, Methods for the calibration of vibration and shock transducers - Part 11: Primary vibration calibration by laser interferometry}
\bibitem{COPA} Ingerslev H, Andresen S and Winther J H 2020
	\newblock{Digital signal processing functions for ultra-low frequency calibrations}, {\it Acta IMEKO}, {\bf 9}, 5, 374
\bibitem{calibration.VIS} Shimoda T, Kokuyama W and Nozato H 2020
	\newblock{A low-acceleration measurement using anti-vibration table with low-frequency resonance}, {\it Acta IMEKO}, {\bf 9}, 5, 369
\end{thebibliography}
\end{document}